\documentclass[aps,prb,twocolumn,floatfix,showpacs]{revtex4}
\usepackage[dvips]{graphicx}
\usepackage{array}
\usepackage{amssymb,amsmath}

\def\ghp{g_{3\|}^{ab}}
\def\ghma{g_{3\perp}^{a}}
\def\ghmb{g_{3\perp}^{b}}
\def\ghmabab{g_{3\perp}^{abab}}
\def\ghmabba{g_{3\perp}^{abba}}
\def\ghmct{g_{3\perp}^{\rm ct}}

\def\gema{g_{1\perp}^{a}}
\def\gemb{g_{1\perp}^{b}}
\def\gemab{g_{1\perp}^{ab}}
\def\gemct{g_{1\perp}^{\rm ct}}

\def\gkma{g_{2\perp}^{a}}
\def\gkmb{g_{2\perp}^{b}}
\def\gkmab{g_{2\perp}^{ab}}
\def\gkmct{g_{2\perp}^{\rm ct}}

\def\gpa{g_\parallel^{a}}
\def\gpb{g_\parallel^{b}}
\def\gpab{g_\parallel^{ab}}
\def\gpct{g_\parallel^{\rm ct}}

\def\gkmaast{g_{2\perp}^{a\ast}}
\def\gkmbast{g_{2\perp}^{b\ast}}
\def\gkmabast{g_{2\perp}^{ab\ast}}

\def\gpaast{g_\parallel^{a\ast}}
\def\gpbast{g_\parallel^{b\ast}}
\def\gpabast{g_\parallel^{ab\ast}}

\def\ghpn{(g_{3\parallel}^{ab})^2}
\def\ghman{(g_{3\perp}^{a})^2}
\def\ghmbn{(g_{3\perp}^{b})^2}

\def\ghmctn{(g_{3\perp}^{\rm ct})^2}
\def\geman{(g_{1\perp}^{a})^2}
\def\gembn{(g_{1\perp}^{b})^2}

\def\gemctn{(g_{1\perp}^{\rm ct})^2}

\def\gkmctn{(g_{2\perp}^{\rm ct})^2}

\def\gpctn{(g_\parallel^{\rm ct})^2}

\def\ka{k_a}
\def\kb{k_b}
\def\va{v_a}
\def\vb{v_b}
\def\d{\textrm{d}}

\begin{document}
\allowdisplaybreaks
\title{Possible phases of two coupled $n$-component fermionic chains}

\author{E. Szirmai and J. S{\'o}lyom}
\affiliation{Research Institute for Solid State Physics and Optics, 
       H-1525 Budapest, P.O.Box 49, Hungary}
\date\today

\begin{abstract}
  A two-leg ladder with $n$-component fermionic fields in the chains has been
  considered using an analytic renormalization group method. The fixed points
  and possible phases have been determined for generic filling as well as for
  a half-filled system and for the case when one of the subbands is half
  filled. A weak-coupling Luttinger-liquid phase and several strong-coupling
  gapped phases have been found. In the Luttinger liquid phase, for the most
  general spin dependence of the couplings, all $2n$ modes have different
  velocities if the interband scattering processes are scaled out, while $n$
  doubly degenerate modes appear if the interband scattering processes remain
  finite. The role of backward-scattering, charge-transfer and umklapp
  processes has been analysed using their bosonic form and the possible phases
  are characterized by the number of gapless modes. As a special case the
  SU($n$) symmetric Hubbard ladder has been investigated numerically. It was
  found that this model does not scale to the Luttinger liquid fixed
  point. Even for generic filling gaps open up in the spectrum of the spin or
  charge modes, and the system is always insulator in the presence of umklapp
  processes.
\end{abstract}

\maketitle

\section{Introduction}

Recently, systems exhibiting non-Fermi-liquid behavior have been intensively 
studied, especially in connection with the unusual normal-state properties of 
high-temperature superconductors. Low-dimensional models are of special 
interest in this respect due to their possible Luttinger-liquid\cite{FDMH} 
behavior. The simplest model that has a non-Fermi-liquid ground state is the 
exactly solvable one-dimensional (1D) Hubbard chain.\cite{EHLFYW} Unless 
the band is half filled, this system is a prototype of Luttinger liquids.

In most strongly correlated systems, the degenerate $d$ or $f$ bands of the
transition-metal or rare-earth ions play an important role and the orbital
degrees of freedom, too, have to be taken into account. In an obvious although
highly simplified generalization of the Hubbard model, it is assumed that the
spin and orbital degrees of freedom can be treated as a unified degree of
freedom with $n$ possible values. The SU($n$) symmetric generalization of the
SU(2) symmetric 1D Hubbard model has been studied by several
authors\cite{RAPAMCPL,JBMIA,ESZJS} to check its possible Luttinger-liquid
behavior and to study the metal-insulator transition at half filling or $1/n$
filling. It has been found that the behavior is qualitatively different for $n
> 2$ and $n=2$.

The physics is more complicated when---as a first step towards two-dimensional
systems---two Hubbard chains or equivalently two Luttinger liquids are coupled
by one-particle hopping into a ladder. The first question that has been
addressed is the relevance or irrelevance of the interchain hopping. It has
been shown\cite{CCCDCWM,MF,DBCBAMST} that for generic filling the
1D Luttinger liquid is unstable with respect to arbitrarily weak
transverse single-particle hopping. Either the system becomes Fermi liquid or
new types of states may occur. The situation may be different at a special
filling where umklapp processes may suppress the single-electron interchain 
hopping.\cite{KLH}

A more difficult problem is to decide what these states may be. To answer 
this question the Hubbard ladder---or more general models with relevant 
intrachain backward scattering processes, too---have been studied by different 
analytic and numerical methods like renormalization group in 
boson\cite{MF,KLH,DVKTMR,HJS,LBMPAF-2,HHLLBMPAF,CWWVLEF} or 
fermion\cite{GAJCNMH,KPJS} representation, Monte Carlo 
simulations\cite{monte-carlo,monte-carlo-2} or using the density-matrix 
renormalization group.\cite{DJS, RMNSRWDJS,YPSLTKL,USSCJOFJBMMT}

As a possible classification of the new states Balents and
Fisher\cite{LBMPAF-2} proposed to characterize the phases by the number of
gapless charge and spin modes. Since a two-leg ladder with two-component
fermions on both legs has two charge and two spin modes, in principle nine
phases can be distinguished in this way. They have shown that seven of them
may appear in the weak-coupling regime of the Hubbard ladder. The others can
be realized in more general models only, which are defined by more coupling
constants.
 
A more complete characterization of the phases can be obtained by
investigating the appropriate correlation functions. When only forward
scattering processes are allowed for on the chains, for generic filling, the
dominant singularity appears in the spin-density and charge-density
responses,\cite{HJS} while if backward-scattering processes are also taken
into account the two-chain model has predominant $d$-type pairing fluctuations
even for purely repulsive interactions.\cite{DVKTMR,HJS} Similar result, the
possibility of density wave or superconductivity has been found in a
variational calculation.\cite{AVR}

Similar situation occurs for the half-filled Hubbard ladder. This model has
been shown\cite{HHLLBMPAF} to scale to the SO(8) Gross-Neveu model which has
fully gapped low-lying excitation spectrum. For repulsive interaction the
system is a Mott-insulator spin liquid with $d$-wave pairing correlations.  It
is interesting to note that if both repulsive and attractive interactions are
allowed for, the phase diagram contains four phases related by SO(5) symmetry,
which can be interpreted\cite{SCZ} as unifying magnetism and
superconductivity. Similar approximate SO(5) symmetry has been
found\cite{DGSDS} at low energies in the repulsive strong-coupling regime of
two coupled Tomonaga-Luttinger chains.

The competition of different density-wave or superconducting orders has been 
studied\cite{CWWVLEF,MTAF} in a more general model, where besides 
the one-particle transverse hopping the Coulomb interaction and Heisenberg 
exchange between sites on the same rungs and along the diagonals of the 
elementrary plaquette as well have been taken into account.

Although the problem of two coupled two-component fermionic chains has been
extensively studied, and some results on the two-dimensional SU($n$) Hubbard
model are known,\cite{JBMIA,CHWH} to the best of our knowledge the 
properties of two coupled SU($n$) symmetric fermionic chains have not been 
discussed. For this reason, in this paper, we will present the results obtained for 
a ladder built up from two $n$-component fermion systems.  In this respect this 
calculaton is an extension of our earlier work.\cite{ESZJS} On the other hand, 
since a two-leg ladder can be treated as a two-band model, and the method 
applied is the multiplicative renormalization group combined with bosonization 
treatment, this paper can also be considered as a generalization of Refs.\ 
\onlinecite{KPJS} and \onlinecite{LBMPAF-2} to $n$-valued spin. 

The model is introduced and the types of interactions are discussed
in Sec.\ \ref{sec:model}. The next section contains the scaling equations.  
The behavior of the system in the weak-coupling Tomonaga-Luttinger fixed point 
and the properties of the gapless Tomonaga-Luttinger phase of the two-band 
model are analyzed in Sec.\ IV. Here the calculations are extended to fully
anisotropic spin-dependent couplings. The role of the backward, charge-transfer and 
umklapp scattering processes is discussed in Sec.\ V, and the possible gapped 
phases characterized by the number of gapped modes are presented. The 
numerical results obtained for the Hubbard ladder and for some more general 
models are presented in Sec.\ VI. Finally Sec.\ VII contains a summary of the 
main results.

\section{The model}
\label{sec:model}

The simplest model we consider is a two-leg SU($n$) Hubbard ladder in which two
identical SU($n$) Hubbard chains described by the Hamiltonians
\begin{equation} \begin{split} \mathcal{H}_{j} & = t
    \sum_{i=1}^N\sum_{\sigma=1}^n \big(
    a_{j,i,\sigma}^{\dagger}a^{\phantom\dagger}_{j,i+1,\sigma} + {\rm
      H.c.}  \big) \\
   \label{eq:H-chain} 
   & \phantom{= ,} + \frac{U}{2} \sum_{i=1}^N\sum_{\sigma, \sigma'=1}^n
   n_{j,i,\sigma}n_{j,i,\sigma'} ,
\end{split} \end{equation} with $j=1,2$, are coupled into a ladder by
a rung hopping term
\begin{equation}
       {\mathcal H}_{\perp} = t_{\perp} \sum_{i=1}^{N} \sum_{\sigma=1}^n \big(
         a_{1,i,\sigma}^{\dagger}a^{\phantom \dagger}_{2,i,\sigma} +
         a_{2,i,\sigma} ^{\dagger}a^{\phantom\dagger}_{1,i,\sigma} \big) .
\label{eq:hopping}
\end{equation}
The local U($n$) symmetry of the interaction term is reduced to global U($n$)
symmetry by the kinetic energy term. An SU($n$) symmetric model is obtained if
the overall U(1) phase factor is removed. Although in real space, the
Hamiltonian has a particularly simple form, in what follows it will be treated
in momentum representation, where it is easier to separate the relevant,
irrelevant or marginal components.

\subsection{Kinetic energy}

\begin{figure}[htb]
  \includegraphics[scale=0.35]{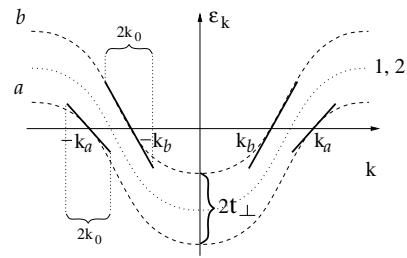}
  \caption{The spectrum of two coupled chains: the spectrum before (after) 
    hybridization is shown by dotted (dashed) lines. Solid lines indicate the 
    linearized spectrum.}
  \label{fig:spectrum1}
\end{figure}

The terms describing the intrachain hopping are diagonal in momentum
representation.  When the interchain hopping is taken into account, the full
kinetic energy is diagonalized by the symmetric and antisymmetric combinations
of the operators belonging to the two chains:
\begin{equation}
\begin{split}
  \label{eq:Bogtrf}
  a_{k,\sigma}\equiv & \, \frac{1}{\sqrt{2}}\big( a_{1,k,\sigma}
           + a_{2,k,\sigma} \big) , \\ 
  b_{k,\sigma}\equiv & \, \frac{1}{\sqrt{2}}\big( a_{1,k,\sigma} 
         - a_{2,k,\sigma} \big) .
\end{split}
\end{equation}
As it is shown in Fig.\ \ref{fig:spectrum1}, the spectrum consists of two
bands labelled by $a$ and $b$ which are obtained by shifting the dispersion
curve of the free chain by $ \pm t_{\perp}$.

\begin{figure}[htb]
  \centering
  \includegraphics[scale=0.35]{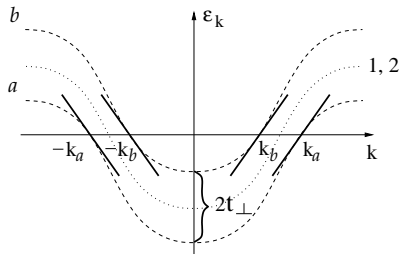}
  \caption{The linearized spectrum for a half-filled ladder.}
  \label{fig:spectrum2}
\end{figure}

When the number of electrons is small, the upper band is completely empty, and
the behavior is identical to that of a one-band Hubbard model. The behavior is
similar in the case when the system is close to saturation, and the lower band
is completely filled. In what follows it will be assumed that the band filling
is between these two extremes, both bands are partially filled, moreover, the
Fermi momenta are not too close to the zone center or zone boundary.  Under
these circumstances one can assume---as is done usually---that the relevant
electronic states are the ones near the Fermi points and their dispersion
relation can be approximated by straight lines with slopes $\pm \hbar \va$ and
$ \pm \hbar \vb$ around the Fermi points $\pm \ka$ and $\pm \kb$,
respectively. If the lattice constant is taken to be unity
\begin{equation}
       \hbar v_{a(b)} = 2 t  \sin k_{a(b)}\,,
\end{equation}
and the value of the Fermi momenta is determined by the overall filling of the
band and the splitting of the subbands, $2t_{\bot}$. It follows from the
symmetry of the tight-binding dispersion relation of the free chain that---as
shown in Fig.\ \ref{fig:spectrum2}---for a half-filled system $\va=\vb$, while
more generally $\va\neq\vb$.

Measuring the momenta from the respective Fermi points the
kinetic energy takes the form
\begin{multline}
  \label{eq:Hkin-2}
  {\mathcal H}_{\textrm{kin}} = \sum_{k,\sigma} \Big[ \hbar \va k \big(
      a_{\ka+k,\sigma} ^\dagger a^{\phantom\dagger}_{\ka+k,\sigma} -
      a_{-\ka+k,\sigma}^\dagger a^{\phantom\dagger}_{-\ka+k,\sigma}\big) \\ +
      \hbar \vb k \big( b_{\kb+k,\sigma}^\dagger
      b^{\phantom\dagger}_{\kb+k,\sigma} - b_{-\kb+k,\sigma}^\dagger
      b^{\phantom\dagger}_{-\kb+k,\sigma}\big) \Big].
\end{multline}
The operators $a_{\pm\ka+k,\sigma}^\dagger$ 
($a^{\phantom \dagger}_{\pm\ka+k,\sigma}$) and $b_{\pm\kb+k,\sigma}^\dagger$ 
($b^{\phantom \dagger}_{\pm\kb+k,\sigma}$) create
(annihilate) an electron with momentum $k$ and spin $\sigma$ around
the Fermi points $\pm\ka$ and $\pm\kb$, respectively. Thus, the
linearized spectrum consists of four branches, two corresponding to
right-moving particles and two to left movers. In what follows the notation
\begin{equation}   \begin{split}
     a^{\phantom \dagger}_{+, k, \sigma} & \equiv 
     a^{\phantom \dagger}_{\ka+k,\sigma} \,, \quad  
       a^{\phantom \dagger}_{-, k, \sigma} \equiv 
      a^{\phantom \dagger}_{-\ka+k,\sigma} \,,  \\ 
     b^{\phantom \dagger}_{+, k, \sigma} & \equiv 
     b^{\phantom \dagger}_{\kb+k,\sigma} \,, \quad  \,
       b^{\phantom \dagger}_{-, k, \sigma} \equiv 
      b^{\phantom \dagger}_{-\kb+k,\sigma}   
\end{split}    \end{equation}
will be used. 

For simplicity a sharp cut-off is assumed in the allowed momenta, $|k|<k_0$,
leading to bandwidths $2E_{a(b)} = 2 \hbar v_{a(b)}k_0$, which in general are
different. It is also assumed that the bandwidth is comparable or less than
the splitting of the bands, in other words the difference $\ka - \kb$ is
comparable or larger than the cut-off $k_0$. Due to this choice, from the
results presented in this paper, it is not possible to recover the limit of
uncoupled chains.

\subsection{Interactions}

When the on-site Hubbard interaction is written in terms of the
operators of right and left-moving fermions of type $a$ and $b$, the
terms appearing have the generic form
\begin{equation}
          \alpha^{\dagger}_{\lambda_1, k_1, \sigma} 
          \beta^{\dagger}_{\lambda_2, k_2, \sigma'} 
          \gamma^{\phantom \dagger}_{\lambda_3, k_3, \sigma'} 
          \delta^{\phantom \dagger}_{\lambda_4, k_4, \sigma} \,,
\label{eq:gen-term}
\end{equation}
where the operators $\alpha^{\dagger}$, $\beta^{\dagger}$, $\gamma$ and
$\delta$ are creation and annihilation operators of particles of type $a$ or
$b$, and $\lambda_i = \pm $. While in the Hubbard ladder the strength of all 
these processes is related to the single on-site Coulomb repulsion $U$, in 
what follows a more general Hamiltonian will be used allowing for different 
couplings for the different scattering processes. 

The coupling constant of the process given in (\ref{eq:gen-term})
will be assumed to be momentum independent and will be denoted by
\begin{equation}
  g^{\alpha\beta\gamma\delta}_{\{\lambda_i\},\sigma \sigma'}  \,.
\end{equation}
It is clear that the terms with couplings
\begin{equation}
   g^{\alpha\beta\gamma\delta}_{\lambda_1,\lambda_2,\lambda_3,\lambda_4,\sigma
    \sigma'} \quad \textrm{and} \quad
    g^{\beta\alpha\delta\gamma}_{\lambda_2,\lambda_1,
    \lambda_4,\lambda_3,\sigma' \sigma}
\label{eq:coupl-iden}
\end{equation}
describe the same process. Similarly, the terms with couplings
\begin{equation}
   g^{\alpha\beta\gamma\delta}_{\lambda_1,\lambda_2,\lambda_3,\lambda_4,\sigma
    \sigma'} \quad \textrm{and} \quad
    g^{\delta\gamma\beta\alpha}_{\lambda_4,\lambda_3,\lambda_2,\lambda_1,
    \sigma \sigma'}
\end{equation}
are Hermitian conjugates to each other. Assuming that the couplings are real,
the same coupling constant is used for them. If the system has inversion
symmetry, the couplings
\begin{equation}
   g^{\alpha\beta\gamma\delta}_{\lambda_1,\lambda_2,\lambda_3,\lambda_4,\sigma
    \sigma'} \quad \textrm{and} \quad
    g^{\alpha\beta\gamma\delta}_{-\lambda_1,-\lambda_2,
    -\lambda_3,-\lambda_4,\sigma \sigma'} ,
\end{equation}
or when combining with (\ref{eq:coupl-iden}) the couplings
\begin{equation}
   g^{\alpha\beta\gamma\delta}_{\lambda_1,\lambda_2,\lambda_3,\lambda_4,\sigma
    \sigma'} \quad \textrm{and} \quad
    g^{\beta\alpha\delta\gamma}_{-\lambda_2,-\lambda_1,
    -\lambda_4,-\lambda_3,\sigma' \sigma} 
\end{equation}
should be taken to be identical. Furthermore, since we have taken the
symmetric and antisymmetric combinations of the operators belonging to the
chains, only such scattering processes survive in which all four particles are
in band $a$ or in band $b$, or two of them are of type $a$ and two of them are
of type $b$.

A scattering process is {\it intraband} if all participating particles belong
to the same subband ($a$ or $b$), $\alpha = \beta=\gamma=\delta$. The
processes in which either $\alpha \neq \beta$ or $\alpha \neq \delta$ are {\it
interband}. The terms with coupling of type $g^{aabb}$ describe processes in
which charge is transferred from one band to the other. If $\sigma \neq
\sigma'$ in an interband process with coupling $g^{abab}$, the scattering is
accompanied by a spin flip in the bands. This could be considered as the
analogue of the Kondo coupling.

It can be assumed that in a non-polarized system the coupling constants depend
on the relative orientation of the spins only. As a further constraint we
will distinguish two cases only, whether $\sigma = \sigma'$ or $\sigma \neq
\sigma'$.  The index $\|$ or $\perp$ will be used for them.

Momentum conservation gives two further restrictions:
\begin{equation}
       k_1 + k_2 = k_3 + k_4   
\end{equation}
and
\begin{equation}
      \lambda_1 k_{\alpha} + \lambda_2 k_{\beta} = \lambda_3 k_{\gamma} + 
        \lambda_4 k_{\delta}  + G \,,
\end{equation} 
where $G$ is a reciprocal lattice vector. $G=0$ corresponds to normal
processes, while $G$ is finite for um\-klapp processes. Since it was assumed
that $\ka$ and $\kb$ are sufficiently different, momentum conservation
(including umklapp processes) can be satisfied only if:
\begin{enumerate}
\item All four particles participating in a scattering process are right
movers or all of them are left movers,
\begin{equation}
   \lambda_1 = \lambda_2 = \lambda_3 = \lambda_4 \,.
\end{equation}
Conventionally these processes are labelled by an index 4. They will be
neglected since---as opposed to the other processes---their contribution is 
not logarithmically divergent in perturbation theory. Normally they lead to 
a Fermi velocity renormalization only. 

\item  Two right (left) movers are scattered into two left (right) movers,
\begin{equation}
    \lambda_1 = \lambda_2 = -\lambda_3 = -\lambda_4 \,.
\end{equation}
These are the umklapp processes and they will be labelled by an index 3.
 
\item A right and a left mover scatter on each other and again a right and
left mover are created. Depending on whether the momentum transfer between the
particles of the same spin is small
\begin{equation}
    \lambda_1 = -\lambda_2 = -\lambda_3 = \lambda_4 
\end{equation}
or large, 
\begin{equation}
    \lambda_1 = - \lambda_2 = \lambda_3 = -\lambda_4 
\end{equation}
the processes are labelled by indices 2 or 1.
\end{enumerate}

\begin{figure}[htb] 
  \centering 
  \includegraphics[scale=0.28]{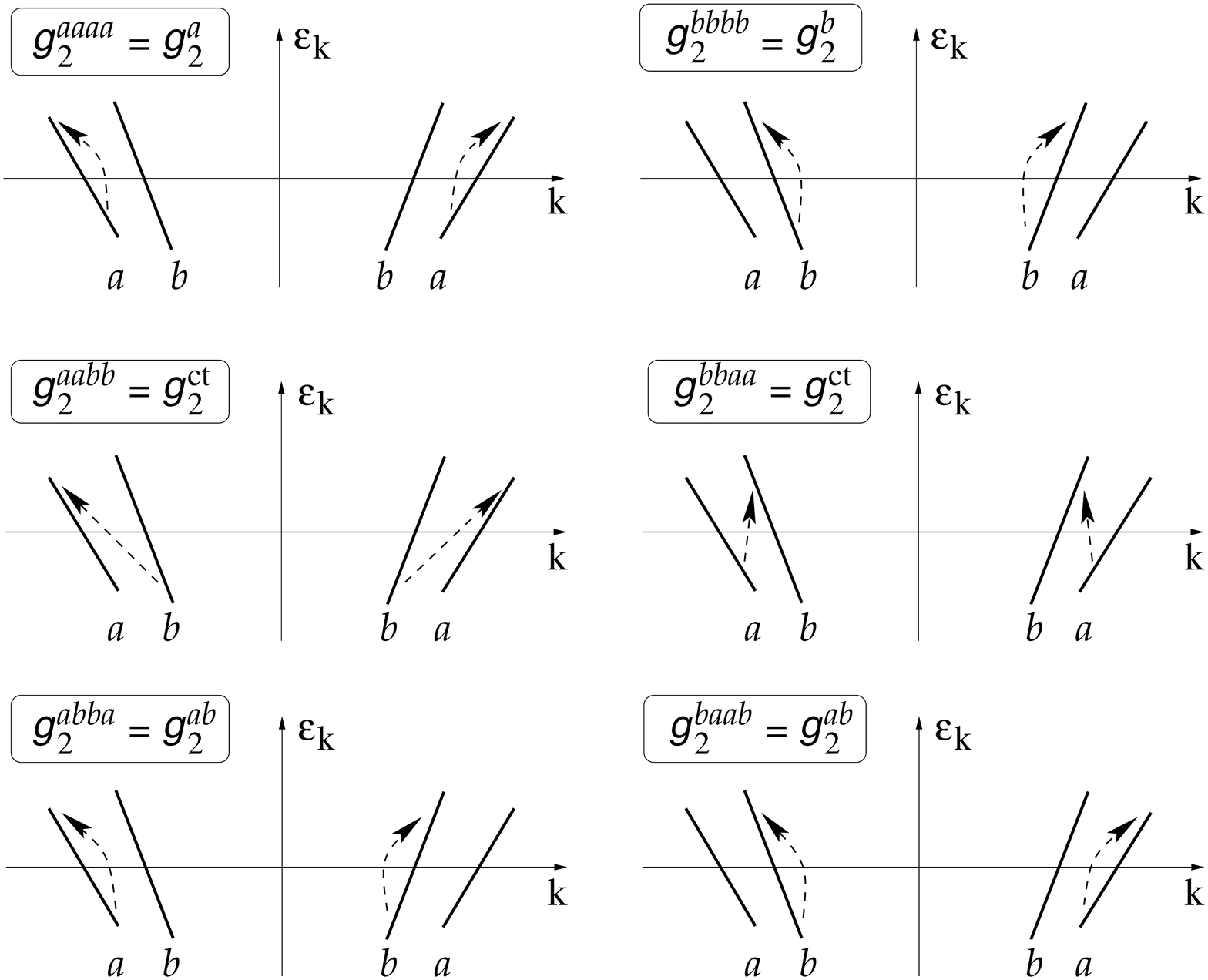}
  \caption{Forward ($g_2$) scattering processes in a ladder system.}
  \label{fig:forward}
\end{figure}

The momentum transfer is small, the process is forward scattering, if either 
$k_{\alpha} - k_{\delta}=0$ (and consequently $k_{\gamma} - k_{\beta}=0$) 
or $ k_{\alpha} - k_{\delta}= \pm (k_a - k_b)$. As shown in Fig.\ 
\ref{fig:forward} six such processes are possible.

Besides the intraband processes, either in band $a$ or in band $b$, there are
four possible interband processes. Two particles from band $b$ are scattered
into band $a$, or vice versa, in channels $a^{\dagger}a^{\dagger}bb$ and
$b^{\dagger}b^{\dagger}aa$, and a small momentum of order $\pm(k_a - k_b)$ is
transferred between right-moving and left-moving particles, respectively.  The
transferred momentum can be arbitrary small in the processes
$a^{\dagger}b^{\dagger}ba$ and $b^{\dagger}a^{\dagger}ab$.  An
$a^{\dagger}b^{\dagger}ab$ process with small momentum transfer is possible
only when all particles are right or left movers, and these processes---as
mentioned above---will be neglected.

Taking into account that the processes $a^{\dagger}a^{\dagger}bb$ and
$b^{\dagger}b^{\dagger}aa$ are Hermitian conjugates and the process 
$b^{\dagger}a^{\dagger}ab$ can be obtained from an $a^{\dagger}b^{\dagger}ba$ 
process by inversion, the
forward-scattering processes are characterized by 8 coupling constants:
\begin{equation}    \begin{split}  
    g^{a}_{2\|} & \equiv g^{aaaa}_{2\|},  \qquad   g^{a}_{2\bot} \equiv 
      g^{aaaa}_{2\bot}, \\
    g^{b}_{2\|} & \equiv g^{bbbb}_{2\|}, \, \qquad  g^{b}_{2\bot} \equiv 
      g^{bbbb}_{2\bot},  \\
    g^{ab}_{2\|} & \equiv g^{abba}_{2\|},  \qquad g^{ab}_{2\bot} \equiv 
     g^{abba}_{2\bot},  \\
    g^{\rm ct}_{2\|}& \equiv g^{aabb}_{2\|},  \qquad g^{\rm ct}_{2\bot} \equiv 
     g^{aabb}_{2\bot}.
\end{split}   \end{equation} 
For six of these eight forward-scattering processes, the particles remain in
the same branch of the spectrum. As we will see, they play a special role. Due
to the conservation of the number of particles and of the spin in each branch,
if they act alone, Luttinger-liquid-like behavior is found.

\begin{figure}[htb] 
  \centering 
  \includegraphics[scale=0.28]{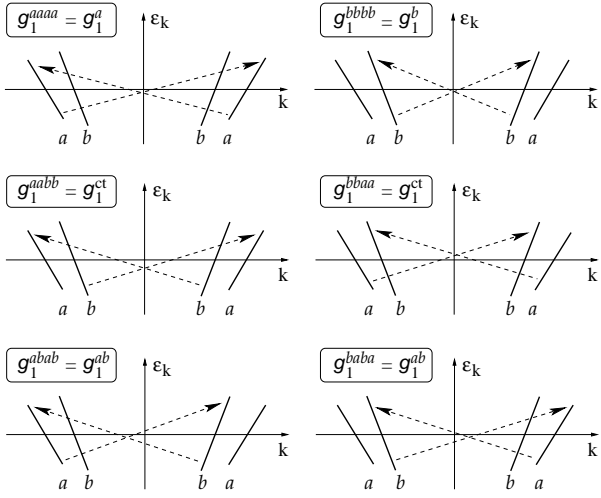}
  \caption{Backward ($g_1$) scattering processes in a ladder system.}
  \label{fig:g12def}
\end{figure}

In the large-momentum-transfer processes, $\lambda_1 = - \lambda_4$ and
$\lambda_2 = - \lambda_3$, and consequently $k_{\alpha} - k_{\delta}= \pm
2k_a$, $\pm 2k_b$ or $\pm (k_a + k_b)$, one has to distinguish normal
(labelled by 1) and umklapp (labelled by 3) processes. Considering first the
normal backward-scattering processes, two of them ($a^{\dagger}a^{\dagger}aa$
and $b^{\dagger}b^{\dagger}bb$) correspond to intraband processes, and four of
them ($a^{\dagger}a^{\dagger}bb$, $b^{\dagger}b^{\dagger}aa$,
$a^{\dagger}b^{\dagger}ab$, and $b^{\dagger}a^{\dagger}ba$) are interband
processes.  The processes $a^{\dagger}b^{\dagger}ba$ and
$b^{\dagger}a^{\dagger}ab$ are forbidden by momentum conservation for finite
hopping amplitude. The six allowed normal backward-scattering processes are
shown in Fig.\ \ref{fig:g12def}.

Since the processes $a^{\dagger}a^{\dagger}bb$ and $b^{\dagger}b^{\dagger}aa$
are Hermitian conjugates to each other and the processes
$a^{\dagger}b^{\dagger}ab$ and $b^{\dagger}a^{\dagger}ba$ are related by
inversion, these processes---analogously to the forward-scattering
ones---could be characterized by 8 coupling constants.  The backward and
forward-scattering terms are, however, not independent when the particles have
identical spins. Looking at the terms with couplings $g_{1\parallel}^{abab}$
and $g_{2\parallel}^{abba}$ given by
\begin{equation}
\begin{split}
  \frac{1}{L}\sum_{k, k', q} \big(g_{1\parallel}^{abab} a_{+, k+q,
    \sigma}^\dagger b_{-, k'-q, \sigma}^\dagger a^{\phantom \dagger}_{+, k',
    \sigma} b^{\phantom \dagger}_{-, k, \sigma} \\ + \, g_{2\parallel}^{abba}
    a_{+, k+q, \sigma}^\dagger b_{-, k'-q, \sigma}^\dagger b^{\phantom
    \dagger}_{-, k', \sigma} a^{\phantom \dagger}_{+, k, \sigma}\big) \,,
\end{split}
\end{equation}
these are the processes in the last rows of Figs.\ \ref{fig:forward} and
\ref{fig:g12def}, it is easily seen that---due to the indistinguishability of
the fermions---they describe exactly the same process, if the model is defined
by a bandwidth cut-off, as is the case in our model. There is only a sign
difference in their contribution. Similar arguments are valid for the
couplings $g_{1\parallel}^{a}$ and $g_{2\parallel}^{a}$ and also for
$g_{1\parallel}^{\rm ct}$ and $g_{2\parallel}^{\rm ct}$. For this reason we
use the combinations:
\begin{equation}
\begin{split}
    g_\parallel^{a}\equiv & \, g_{1\parallel}^{a}-g_{2\parallel}^{a}, \\
   g_\parallel^{b}\equiv & \, g_{1\parallel}^{b}-g_{2\parallel}^{b}, \\
   g_\parallel^{ab} \equiv & \, g_{1\parallel}^{ab}-g_{2\parallel}^{ab}, \\
   g_\parallel^{\rm ct} \equiv & \, g_{1\parallel}^{\rm
   ct}-g_{2\parallel}^{\rm ct},
\end{split}
\end{equation}
and only the processes with couplings $g_{1 \perp}^{a}$, $g_{1\perp}^{b}$,
$g_{1 \perp}^{ab}$, and $g_{1 \perp}^{\rm ct}$ are true backward-scattering
processes.

\begin{figure}[htb] 
  \centering 
  \includegraphics[scale=0.28]{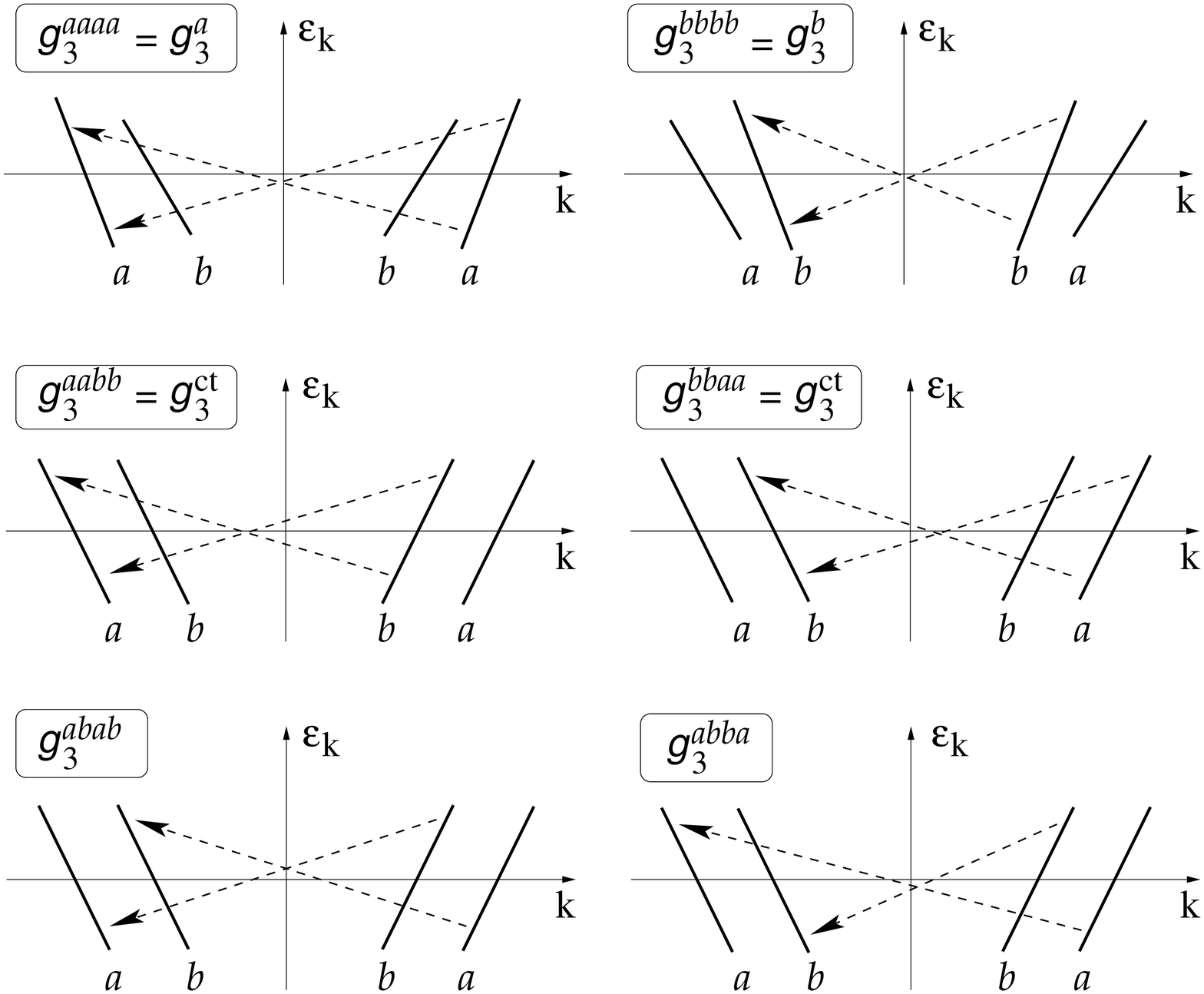}
  \caption{Umklapp ($g_3$) processes in a ladder system.}
  \label{fig:g3def}
\end{figure}

When either band $a$, band $b$ or the whole spectrum is half-filled, it is
possible that $\lambda_1 k_{\alpha} + \lambda_2 k_{\beta} = \lambda_3
k_{\gamma} + \lambda_4 k_{\delta} + G$ with a vector of the reciprocal
lattice. In the first case intraband umklapp processes are allowed for, while
in the second case there are four kinds of interband umklapp processes as
shown in Fig.\ \ref{fig:g3def}, but as mentioned above, two of them appear
with the same coupling constant.

Due to the requirement of antisymmetry of the products of fermion operators,
no intraband umklapp processes are allowed between particles of the same
spin. This is also true for the charge-transfer umklapp processes of type
$a^{\dagger}_{+,\sigma}a^{\dagger}_{+,\sigma}b^{\phantom\dagger}_{-,\sigma}
b^{\phantom\dagger}_{-,\sigma}$. However, interband umklapp processes of type
$a^{\dagger}_{+,\sigma}b^{\dagger}_{+,\sigma}a^{\phantom\dagger}_{-,\sigma}
b^{\phantom\dagger}_{-,\sigma}$ and
$a^{\dagger}_{+,\sigma}b^{\dagger}_{+,\sigma}
b^{\phantom\dagger}_{-,\sigma}a^{\phantom\dagger}_{-,\sigma}$ are not excluded
by symmetry. Since they correspond to the same type of process the
notation
\begin{equation}
g_{3\parallel}^{ab}\equiv g_{3\parallel}^{abab}-g_{3\parallel}^{abba}
\end{equation}
will be used. Thus altogether two intraband and four interband umklapp
processes with different coupling constants could be distinguished.

In order to get simpler formulas in what follows the dimensionless couplings
\begin{equation}
  g_i^{\alpha \beta \gamma \delta} \equiv \frac{g_i^{\alpha \beta 
      \gamma \delta}}{\frac{1}{2}\pi \hbar (v_{\alpha} + v_{\beta} 
    + v_{\gamma} + v_{\delta}) }
\label{eq:coupl-dimless}
\end{equation}
and the quantity
\begin{equation}
  \gamma=\frac{(\va+\vb)^2}{4\va\vb}
\end{equation}
will be used. $\gamma$ characterizes the relative slopes of the two
bands. $\gamma=1$ if and only if $\va=\vb$, while $\gamma>1$ otherwise, 
if both $\va$ and $\vb$ are positive. 

\section{Renormalization-group treatment}
\label{sec:sceq}

The perturbative corrections to the vertices and self energy of the
1D interacting fermion system defined by the Hamiltonian in
(\ref{eq:H-chain}) and (\ref{eq:hopping}) are logarithmically singular. A
convenient procedure to sum up the leading or next-to-leading logarithmic
corrections to any order of the couplings is provided by the multiplicative
renormalization-group method (for a detailed description see
Ref.\ \onlinecite{JS}). Although the bandwidth is different in the two bands,
in the logarithmic approximation a common cut-off can be used,\cite{KPJS}
i.e., in the argument of the logarithm, a quantity $\omega/E_0$ will be used 
instead of $\omega/E_{a}$, $\omega/E_{b}$ or $2\omega/(E_{a}+ E_{b})$.

Up to second order in the couplings the renor\-ma\-li\-za\-tion-group
procedure leads to the following scaling equations (here $x=E_0'/E_0$, and
prime denotes the renormalized cut-off): for the forward-scattering terms
without charge transfer between the bands
\begin{subequations}
\label{eq:se1}
\begin{flalign}
\label{eq:se1-e}
  \frac{\d \gpa}{\d \ln x} =  & \, (n-1)(\gema)^2 - \gamma\gpctn  \\
 & \, + (n-1)\Big[(\ghmabba)^2+\ghman\Big] + \ghpn , \nonumber  \\
  \frac{\d \gkma}{\d \ln x} = & \, (\gema)^2 + \gamma\gemctn + 
  \gamma\gkmctn \\
   & \, - (\ghmabab)^2-\ghman ,   \nonumber  \\
  \frac{\d \gpb}{\d \ln x} =  & \, (n-1)(\gemb)^2 - \gamma\gpctn  \\
 & \, + (n-1)\Big[(\ghmabba)^2+\ghmbn\Big] + \ghpn ,  \nonumber   \\
  \frac{\d \gkmb}{\d \ln x} = & \, (\gemb)^2 + \gamma\gemctn + 
  \gamma\gkmctn \\
   & \, - (\ghmabab)^2-\ghmbn ,   \nonumber  \\
  \frac{\d \gpab}{\d \ln x} = & \, \gpctn + (n-1)\Big[ 
  \gemctn + (\gemab)^2   \\  & \,\,\, + \ghmctn + (\ghmabab)^2 
  \Big] + \ghpn ,   \nonumber  \\
  \frac{\d \gkmab}{\d \ln x} = & \, (\gemab)^2 - \gkmctn -
  (\ghmabba)^2 -\ghmctn ,
\end{flalign}
\end{subequations}
for the backward-scattering terms without charge transfer
\begin{subequations}
\label{eq:be1}
\begin{flalign}
\label{eq:be1-e}
  \frac{\d \gema}{\d \ln x}  = & \, 2\gema\Big[\gpa+\gkma\Big]
  + 2\gamma\gemct\gkmct  - 2\ghp\ghmabba    \\
     & \, + (n-2)\Big[\geman + (\ghmabba)^2+\ghman \Big] , \nonumber \\
   \frac{\d \gemb}{\d \ln x}  = & \, 2\gemb\Big[\gpb+\gkmb\Big]
  + 2\gamma\gemct\gkmct   - 2\ghp\ghmabba  \\
     & \,  + (n-2)\Big[\gembn + (\ghmabba)^2+\ghmbn \Big] , \nonumber \\
 \frac{\d \gemab}{\d \ln x} = & \, 2\gemab\Big[\gpab + \gkmab\Big]  
  + 2\gemct\gpct  +2\ghp\ghmabab    \\
    & \, + (n-2)\Big[ (\gemab)^2   + \gemctn +
  (\ghmabab)^2  + \ghmctn  \nonumber \Big], 
\end{flalign}
\end{subequations}
for the charge-transfer terms 
\begin{subequations}
\label{eq:ct1}
\begin{flalign}
 \frac{\d \gpct}{\d \ln x} = & \,- \gpct\Big[\gpa + \gpb - 2\gpab\Big] \\ &
 \, + 2(n-1)\Big[\gemct\gemab + \ghmct\ghmabab\Big] , \nonumber \\ \frac{\d
 \gkmct}{\d \ln x} = & \, \gemct\Big[\gema + \gemb\Big] \\ & \, +
 \gkmct\Big[\gkma + \gkmb - 2\gkmab\Big] - 2\ghmct\ghmabba , \nonumber \\
 \frac{\d \gemct}{\d \ln x} = & \, \gemct\Big[\gkma + \gkmb + 2\gpab +
 2(n-2)\gemab\Big] \nonumber \\ & \, + \gkmct\Big[\gema + \gemb\Big]  +
 2\gpct\gemab \\ & \, +2(n-2)\ghmct\ghmabab + 2\ghp\ghmct , \nonumber
\end{flalign}
\end{subequations}
for the intraband umklapp processes
\begin{subequations}
\label{eq:uma}
\begin{flalign}
\label{eq:uma-e}
  \frac{\d \ghma}{\d \ln x} = & \, 2\ghma\left[\gpa + (n-2)\gema - \gkma 
  \right] , \\ 
  \frac{\d \ghmb}{\d \ln x} = & \, 2\ghmb\left[\gpb + (n-2)\gemb - \gkmb 
  \right] , 
\end{flalign}
\end{subequations}
and for the interband umklapp processes
\begin{widetext}
\begin{subequations}
\label{eq:um1}
\begin{flalign}
\label{eq:um1-e}
  \frac{\d \ghmabab}{\d \ln x} = & \, \ghmabab\Big[2\gpab + 2(n-2)\gemab -
  \gkma-\gkmb\Big] + 2\ghp\gemab + 2\ghmct\Big[\gpct+ (n-2)\gemct\Big] ,\\
  \frac{\d \ghmabba}{\d \ln x} = & \, \ghmabba\Big[\gpa+\gpb- 2\gkmab+(n-2)
  \left(\gema+\gemb\right)\Big] - 2\ghmct\gkmct -\ghp\Big[\gema +\gemb\Big]
  ,\\
  \label{eq:se1-u}
  \frac{\d \ghp}{\d \ln x} = & \, -(n-1)\Big[2\ghmabab\gemab + 2\ghmct\gemct
  \Big] - \ghp\Big[2\gpab + \gpa + \gpb\Big] + (n-1)\ghmabba\Big[\gema +
  \gemb\Big] , \\ \frac{\d \ghmct}{\d \ln x} = & \, 2\ghmct\left(\gpab +
  (n-2)\gemab- \gkmab\right) + 2\ghmabab\left(\gpct+(n-2)\gemct\right) -
  2\ghmabba\gkmct + 2\ghp\gemct .
\end{flalign}
\end{subequations}
\end{widetext}

These systems of equations are the generalizations of the set of equations 
investigated in detail in Ref.\ \onlinecite{KPJS} to $n$-valued spin and to 
the case when umklapp processes are taken into account. As it will be
discussed, depending on the filling of the bands some or all umklapp processes
could be neglected, because momentum conservation cannot be satisfied for
states near the Fermi points.

Once the scaling equations have been derived one can look for their fixed
points. Two types of fixed points can be distinguished. One of them
corresponds to fixed points, where some of the couplings vanish while all
others remain weak, the dimensionless couplings introduced in
(\ref{eq:coupl-dimless}) are less than unity. In what follows the term
fixed point is used even if the marginal couplings scale to non-universal
values on an extended hypersurface in the space of couplings.

In the other type of fixed points at least some of the couplings scale to
large values. Using the scaling equations derived in the leading logarithmic
approximation these couplings scale to infinity.  When the next-to-leading
logarithmic corrections are taken into account, finite strong-coupling fixed
points are obtained, but they are located outside the region of validity of
the low-order calculation. On the basis of this perturbative approach one
cannot decide whether these couplings scale to a finite fixed point with
dimensionless strength larger than unity or to infinity. Fortunately, for the
characterization of the phases it is sufficient to know whether these
couplings are relevant or irrelevant.

\section{Tomonaga-Luttinger fixed point}
\label{sec:fixp}

In the most general case the model is characterized by 18 coupling constants,
six of them belong to forward-scattering processes without charge transfer,
three are backward-scattering processes without charge transfer, three
correspond to charge-transfer processes and six to umklapp
processes. When all true backward-scattering processes are irrelevant, their
coupling constants scale to zero,
\begin{equation}
  \label{eq:fixp1}
  g_{1\perp}^{a \ast} =g_{1\perp}^{b\ast} = g_{1\perp}^{ab\ast} = 0 \,,
\end{equation}
all processes in which charge is transferred from one band to the other
are also irrelevant,
\begin{equation}
\label{eq:fixp1kieg}
  g_{\|}^{{\rm ct}\ast} = g_{1\perp}^{{\rm ct}\ast} = g_{2\perp}^{{\rm
  ct}\ast} = 0\,,
\end{equation}
and all umklapp processes are irrelevant, the right-hand sides of the scaling
equations for the couplings of the forward-scattering processes vanish
also. This fixed point will be denoted as FP--TL, referring to the fact that
only those processes survive in which charge and spin conservation are
statisfied in each branch of the spectrum, and therefore Tomonaga-Luttinger
behavior is expected.

\subsection{Stability analysis}

The behavior of the couplings close to a fixed point can be studied 
analytically using the eigenvalues of the linearized matrix 
\begin{equation}
\label{eq:matrix}
     L_{ij} =\frac{g_i'}{g_j}\bigg|_{\{g\ast\}}
\end{equation}
of the renormalization-group transformation, where prime denotes the
renormalized coupling, $i$ and $j$ are short-hand notations for all indices,
and $\ast$ refers to the fixed-point value. Deviations from the fixed-point
are relevant if the eigenvalue is larger than unity, they are irrelevant
if the eigenvalue is less than unity. A unit eigenvalue indicates marginal
direction. Whether a marginal direction is marginally relevant or marginally
irrelevant is determined by higher-order corrections.

First the case of generic band filling is studied, when all umklapp processes
can be neglected. In the weak-coupling regime near FP--TL the matrix $L_{ij}$
defined in (\ref{eq:matrix}) can be diagonalized. The 12 eigenvalues are:
\begin{subequations}
\label{eq:ev1}
\begin{flalign}
\label{eq:ev1-e}
\lambda_1 = &\,1+2(\gpaast+\gkmaast)\ln x, \\
\lambda_2 = &\,1+2(\gpbast+\gkmbast)\ln x, \\
\lambda_3 = &\,1+2(\gpabast+\gkmabast)\ln x, \\
\lambda_4 = &\,1+(-\gpaast-\gpbast+2\gpabast)\ln x, \\
\lambda_5 = &\,1+(\gkmaast+\gkmbast+2\gpabast)\ln x, \\
\label{eq:ev1-u}
\lambda_6 = &\,1+(\gkmaast+\gkmbast-2\gkmabast)\ln x 
\end{flalign}
\end{subequations}
and the six other eigenvalues are equal to 1. The fixed point can be
attractive if the eigenvalues $\lambda_1, \dots, \lambda_6$ are not greater
than unity. Since during the renormalization procedure the cut-off is scaled
to smaller values, the logarithmic multiplicative factor has negative sign, the
coupling constant combinations appearing in the brackets have to be
non-negative in the irrelevant directions. Thus for generic band filling 
FP-TL can be attractive if the following inequalities hold:
\begin{subequations}
\label{eq:req}
\begin{flalign}
\label{eq:req-e}
  \gpaast \geq & \, -\gkmaast, \\
  \gpbast \geq & \, -\gkmbast, \\
  \gpabast \geq & \, -\gkmabast, \\
  2\gpabast \geq & \, \gpaast + \gpbast, \\
  2\gpabast \geq & \, -\gkmaast - \gkmbast, \\
\label{eq:req-u}
  \gkmaast + \gkmbast \geq & \, 2\gkmabast.
\end{flalign}
\end{subequations}
Although these relations give the stability condition of the fixed point only
and cannot be used to get the whole basin of attraction, clearly the
Tomonaga-Luttinger fixed point has an extended basin of attraction in the
parameter space.

When either band $a$ or band $b$ is half filled, i.e., $4\ka$ or $4\kb$ is
equal to a reciprocal lattice vector $G$, the contributions of the intraband
umklapp processes have to be taken into account. We will consider the process
$\ghma$ but the results can be easily extended to the case when $\ghmb$ is
relevant by replacing the couplings $g^{a}_i$ by the corresponding
$g^{b}_i$. FP-TL is recovered if besides the requirements given in
(\ref{eq:fixp1}) (backward-scattering processes are irrelevant) and
(\ref{eq:fixp1kieg}) (charge-transfer processes are irrelevant) the intraband
umklapp process, too, is irrelevant, the fixed-point value of its coupling
vanishes:
\begin{equation}
  \label{eq:fixp2}
   g_{3\perp}^{a \ast}=0.
\end{equation}

With one new coupling added the 13 eigenvalues of the stability matrix
$L_{ij}$ are as follows: 12 of them are the same as given earlier in
\eqref{eq:ev1}, and the 13th eigenvalue is
\begin{equation}
\label{eq:ev2}
\lambda_{13} = 1 + 2(\gpaast - \gkmaast)\ln x.
\end{equation}
The condition of stability of the fixed point is similar to the case discussed
above, except that the requirement $\gpaast \geq \gkmaast$ changes the
first inequality to
\begin{equation}
  \gpaast \geq |\gkmaast |.
\end{equation}
This means that Luttinger-liquid behavior is expected in a smaller region of
the parameter space.

The situation is similar when the system is half-filled, $2(\ka+\kb)$ is equal
to a reciprocial lattice vector $G$ and therefore the interband umklapp 
processes have to be taken into account. In this case the two Fermi velocities
are equal ($\gamma=1$). A weak-coupling FP-TL is obtained if besides the 
backward-scattering processes (\ref{eq:fixp1}) and charge-transfer 
processes (\ref{eq:fixp1kieg}) all interband umklapp processes
are irrelevant,
\begin{equation}
  \label{eq:fixp3}
  {g_{3\perp}^{abab}}^* = {g_{3\perp}^{abba}}^* = {g_{3\perp}^{aabb}}^* =
          {g_{3\parallel}^{ab}}^* = 0 \,.
\end{equation}

In this case 16 coupling constants have to be taken into account. 12 of the 16
eigenvalues are the same as for generic filling and the remaining 4 are:
\begin{subequations}
\label{eq:ev3}
\begin{flalign}
\lambda_{13} = &\,1+2(\gpabast-\gkmabast)\ln x, \\
\lambda_{14} = &\,1+(-\gkmaast-\gkmbast+2\gpabast)\ln x, \\
\lambda_{15} = &\,1+(\gpaast+\gpbast-2\gkmabast)\ln x, \\
\lambda_{16} = &\,1+(-\gpaast-\gpbast+2\gpabast)\ln x, 
\end{flalign}
\end{subequations}
The attractive region of the weak-coupling fixed point is determined
by the inequalities:
\begin{subequations}
\begin{flalign}
  \gpaast  \geq & \, -\gkmaast , \\
  \gpbast  \geq & \, -\gkmbast , \\
  \gpabast  \geq & \, |\gkmabast  |, \\
  2\gpabast  \geq & \, \gpaast  + \gpbast , \\
  2\gpabast  \geq & \, |\gkmaast  + \gkmbast| , \\
  \gkmaast  + \gkmbast  \geq & \, 2\gkmabast , \\
  \gpaast  + \gpbast  \geq & \, 2\gkmabast  
\end{flalign}
\end{subequations}
The basin of attraction of the Tomonaga-Luttinger fixed point is further
reduced by the requirement that all interband umklapp processes should be
irrelevant.

One can see that FP-TL cannot be reached when the model is fully attractive,
when the initial values of all couplings are negative. On the other hand in a
fully repulsive model, when the initial values of all couplings are positive,
Luttinger-liquid behavior is a possibility. Unfortunately, as will be seen,
the repulsive Hubbard ladder does not belong to this class.

\subsection{Bosonization of the Tomonaga-Luttinger part of the model}
\label{sec:TL}

It is expected that for such couplings which scale to the weak-coupling
FP-TL, this $n$-component fermionic ladder has $2n$
gapless bosonic excitations, two soft charge and 2$(n-1)$ soft spin modes, and
therefore the model is equivalent to a 2$n$-component Luttinger liquid. This
is shown here by transforming the fixed-point Hamiltonian into bosonic
language. For this a more general model will be considered where only
intraband and interband forward-scattering terms are allowed without charge
transfer, but the couplings $g_{2;\sigma\sigma'}^{a}$,
$g_{2;\sigma\sigma'}^{b}$, and $g_{2;\sigma\sigma'}^{ab}$ may depend on
$\sigma$ and $\sigma'$ separately.  It is assumed, however, that they are
symmetric in the spin indices,
\begin{equation}
        g_{2;\sigma\sigma'}^{a(b,ab)}  = g_{2;\sigma'\sigma}^{a(b,ab)} \,. 
\end{equation}

Introducing the density operators corresponding to particle-hole excitations 
with small momentum,
\begin{equation}  \begin{split}
  n_{\pm,\sigma}^a(q) & = \sum_k  
  a_{\pm,k,\sigma}^\dagger a^{\phantom \dagger}_{\pm,k+q,\sigma}\, ,  \\
  n_{\pm,\sigma}^b(q) & = \sum_k  
  b_{\pm,k,\sigma}^\dagger b^{\phantom \dagger}_{\pm,k+q,\sigma}\, ,
\end{split}    \end{equation}
the kinetic energy of the particles in the two bands given in
\eqref{eq:Hkin-2} can be written as usual in the form
\begin{equation}
  {\mathcal H}_\textrm{kin} = \pi \hbar \sum_{\lambda,\sigma} \int \d
  x \Big[ \va \big(n_{\lambda,\sigma}^a(x)\big)^2 +\vb
  \big(n_{\lambda,\sigma}^b(x)\big)^2 \Big]\, ,
\end{equation}
where $\lambda=\pm$ and the operators $n_{\lambda,\sigma}^{a(b)}(x)$ are the
densities in real space, the Fourier transforms of
$n_{\lambda,\sigma}^{a(b)}(q)$.  The low-energy part of the spectrum contains
excitations with large momentum, too.  They correspond to excitations in which
the number of particles with a given spin index in a given branch
changes. They can be taken into account by giving the change in the number of
particles in the branches. For simplicity we will not write out these terms
since they do not influence the results.

The interaction terms containing intraband and interband forward-scattering
processes without charge transfer can also be expressed in terms of the
small-momentum components of the densities:
\begin{flalign}
  {\mathcal H}^\textrm{Lutt}_\textrm{int} & = \frac{2 \pi \hbar}{L}
  \sum_{\sigma\sigma'} \sum_q \Big[ v_a
  g_{2;\sigma\sigma'}^{a}n_{+,\sigma}^a(q) n_{-,\sigma'}^a(-q) \nonumber \\ &
  \hskip 1cm + v_b g_{2;\sigma\sigma'}^{b} n_{+,\sigma}^b(q)
  n_{-,\sigma'}^b(-q) \\ & \hskip 1cm + {\textstyle\frac{1}{4}}(v_a + v_b)
  g_{2;\sigma\sigma'}^{ab} n_{+,\sigma}^a(q) n_{-,\sigma'}^b(-q) \nonumber \\
  & \hskip 1cm + {\textstyle\frac{1}{4}}(v_a + v_b) g_{2;\sigma\sigma'}^{ab}
  n_{+,\sigma}^b(q) n_{-,\sigma'}^a(-q) \Big], \nonumber
\end{flalign}
which can be written in real-space representation as
\begin{flalign}
{\mathcal H}^\textrm{Lutt}_\textrm{int} & = 2 \pi \hbar \sum_{\sigma\sigma'}
 \int \d x \Big[ v_a g_{2;\sigma\sigma'}^{a}n_{+,\sigma}^a(x)
 n_{-,\sigma'}^a(x) \nonumber \\ & \hskip 1cm + v_b g_{2;\sigma\sigma'}^{b}
 n_{+,\sigma}^b(x) n_{-,\sigma'}^b(x) \\ &\hskip 1cm +
 {\textstyle\frac{1}{4}}(v_a + v_b) g_{2;\sigma\sigma'}^{ab}n_{+,\sigma}^a(x)
 n_{-,\sigma'}^b(x) \nonumber \\ & \hskip 1cm + {\textstyle\frac{1}{4}}(v_a +
 v_b) g_{2;\sigma\sigma'}^{ab} n_{+,\sigma}^b(x) n_{-,\sigma'}^a(x)\Big] .
 \nonumber
\end{flalign}

First, the Hamiltonian is diagonalized in its spin indices $\sigma$ and
$\sigma'$. To this end we perform an orthogonal transformation to a new
basis corresponding to such combinations of the bosonic densities, that two of 
them (one for each band) are symmetric and $2(n-1)$ ($n-1$ for each band) are 
antisymmetric in the spin quantum numbers:
\begin{flalign}
n_{\lambda,{\rm c}}^{a(b)}(x) & = \frac{1}{\sqrt{n}}\sum_{\sigma = 1}^n
        n_{\lambda,\sigma}^{a(b)}(x) \, ,\\ 
n_{\lambda,m{\rm s}}^{a(b)}(x) & =
        \frac{1}{\sqrt{m(m+1)}} \left(\sum_{\sigma=1}^m
        n_{\lambda,\sigma}^{a(b)}(x) - m n_{\lambda,m+1}^{a(b)}(x)\right) ,
        \nonumber
\end{flalign}
where $m=1,\dots,n-1$. This transformation leads to a Hamiltonian
in which the spin and charge degrees of freedom are separated,
\begin{equation}
   {\mathcal H} = \sum_j \int \d x \big(\mathcal{H}_{\textrm{kin},j}(x) 
      + \mathcal{H}_{\textrm{int},j}(x)\big)\,,
\end{equation}
where $j = \textrm{c}, 1\textrm{s}, 2\textrm{s},\dots, (n-1)\textrm{s}$, and
the kinetic energy part of the Hamiltonian density is
\begin{equation}
  \mathcal{H}_{\textrm{kin},j}(x) = \pi \hbar \sum_\lambda
  \Big[\va \big(n_{\lambda,j}^a(x)\big)^2
  +\vb \big(n_{\lambda,j}^b(x)\big)^2 \Big] ,
\end{equation}
while the interaction term can be written as
\begin{flalign}
  \mathcal{H}_{\textrm{int},j}(x) & = 2\pi\hbar\Big( v_a g^{a}_{2;j}
 n_{+,j}^a(x) n_{-,j}^a(x) \nonumber \\ & + v_b g^{b}_{2;j}n_{+,j}^b(x)
 n_{-,j}^b(x) \\ & + {\textstyle \frac{1}{4}}g^{ab}_{2;j}(v_a+v_b)
 n_{+,j}^a(x) n_{-,j}^b(x) \nonumber \\ & + {\textstyle
 \frac{1}{4}}g^{ab}_{2;j}(v_a+v_b)n_{+,j}^b(x) n_{-,j}^a(x)\Big)\nonumber,
\end{flalign}
and the corresponding couplings are
\begin{flalign}
  g_{2;\textrm{c}}^{a(b,ab)} & = \frac{1}{n}\sum_{\sigma \sigma'=1}^{n}
         g_{2;\sigma\sigma'}^{a(b,ab)}\,, \\ g_{2;m\textrm{s}}^{a(b,ab)} & =
         \frac{1}{m(m+1)}\sum_{\sigma \sigma'=1}^m
         g_{2;\sigma\sigma'}^{a(b,ab)} + \frac{m}{m+1} g_{2;m+1,m+1}^{a(b,ab)}
         \nonumber \\ & \phantom{+} - \frac{1}{m+1} \sum_{\sigma=1}^m
         \left(g_{2;\sigma,m+1}^{a(b,ab)} + g_{2;m+1,\sigma}^{a(b,ab)} \right)
         \, .  \nonumber
\end{flalign}

In order to obtain the velocities of the different modes the Hamiltonian is
rewritten in terms of continuum bosonic fields. The fields
$\phi_{\lambda,j}^{a(b)}(x)$ are related to the densities by the relation
\begin{equation}
  \pi n_{\lambda,j}^{a(b)}(x)=-\partial_x\phi_{\lambda,j}^{a(b)}(x).
\end{equation}
The total phase fields and their duals are
\begin{equation}  \begin{split}
  \phi_j^{a(b)}(x) & = \phi_{+,j}^{a(b)}(x)+\phi_{-,j}^{a(b)}(x), \\
  \theta_j^{a(b)}(x) & = \phi_{+,j}^{a(b)}(x)-\phi_{-,j}^{a(b)}(x) \,.
\end{split}  \end{equation}
The fields $\Pi_j^{a(b)}$ defined by 
\begin{equation}
  \pi \Pi_j^{a(b)}(x)=-\partial_x\theta_{j}^{a(b)}(x)
\end{equation}
are canonical conjugates to the fields $\phi_{j}^{a(b)}(x)$, they satisfy 
the commutation relation:
\begin{equation}
\big[\Pi_j^{a(b)}(x),\phi_{j'}^{a(b)}(x')\big]=i \delta_{jj'}\delta(x-x') \,.
\end{equation}

In this representation, the Hamiltonian can be written as
\begin{equation}
   {\mathcal H} = \sum_j \int \d x \big( \mathcal{H}^{a}_{j}(x) +
     \mathcal{H}^{b}_{j}(x) + \mathcal{H}^{ab}_j(x) \big)\,,
\end{equation}
where
\begin{equation}
\mathcal{H}^{a}_j(x) = \frac{\hbar u_{a,j}}{2\pi}
\Big[\pi^2K_{a,j}\big(  \Pi^{a}_j(x)\big)^2 + \frac{1}{K_{a,j}}\big(\partial_x
  \phi_j^a(x)\big)^2\Big],
\end{equation}
with 
\begin{equation}   \begin{split}
  u_{a,j} & = v_a \sqrt{1 - (g_{2;j}^{a})^2 }\, , \\
  K_{a,j} & =\sqrt{\frac{1-g_{2;j}^{a}}{1 + g_{2;j}^{a}}},
\end{split}    \end{equation}
and similar expressions with index $b$, and the coupling between the bands 
is given by
\begin{equation}   \begin{split}
  \mathcal{H}^{ab}_j(x) = \frac{\hbar(v_a+v_b)}{4\pi}g_{2;j}^{ab}
  \Big[ & \partial_x\phi_j^a(x)\partial_x\phi_j^b(x) \\ 
     & - \pi^2\Pi_j^a(x)\Pi_j^b(x)\Big].
\end{split}  \end{equation}

The first two terms of the Hamiltonian can be interpreted as the Hamiltonian
of two $n$-component Luttinger liquids. The coupling $\mathcal{H}^{ab}_j(x)$
is diagonal in the component index, it couples the $2n$ components into $n$
pairs.  Without this coupling, in the most general case, the 2$n$ modes have
different velocities, so not only spin-charge separation occurs in the system
but all modes appear separately. In the special case, when the couplings
depend only on whether $\sigma = \sigma'$ or $\sigma \neq \sigma'$, the
couplings are
\begin{equation}
  g^{a(b)}_{2;\textrm{c}} = g^{a(b)}_{2\|} + (n-1) g^{a(b)}_{2\bot}\,,
\end{equation}
and
\begin{equation}
  g^{a(b)}_{2;m\textrm{s}} = g^{a(b)}_{2\|} -  g^{a(b)}_{2\bot}\,,
  \qquad m=1,2,\dots,(n-1) \,,
\end{equation}
so the $n-1$ spin modes in band $a$ $(b)$ have the same velocity. 

The full Hamiltonian can be diagonalized by the ``symmetric'' and
``antisymmetric'' combinations of the operators belonging to the two bands:
\begin{equation}
  \tilde{\phi}_j^{\pm}  = \frac{1}{\sqrt{u_{a,j}+u_{b,j}}}\Big[\phi_j^a
    \sqrt{u_{a,j}/K_{a,j}}\pm \phi_j^b \sqrt{u_{b,j}/K_{b,j}}\Big] ,
\end{equation}
and similar expression holds for the conjugated momenta:
\begin{equation}
  \tilde{\Pi}_j^{\pm}  = \frac{1}{\sqrt{u_{a,j}+u_{b,j}}}\Big[\Pi_j^a
    \sqrt{u_{a,j}K_{a,j}} \pm  \Pi_j^b \sqrt{u_{b,j}K_{b,j}}\Big],
\end{equation}
such that they satisfy the canonical commutation relation.

With these operators the Hamiltonian density is the sum of $2n$
Tomonaga-Luttinger Hamiltonians:
\begin{equation}
  \mathcal{H}(x) = \sum_j \left( \mathcal{H}^+_j(x) + \mathcal{H}^-_j(x)
  \right)
\end{equation}
where
\begin{equation}
  \mathcal{H}^{\pm}_j(x) = \frac{\hbar \tilde{u}_{\pm,j}}{2\pi}
  \Big[\pi^2\tilde{K}_{\pm,j} \big( \tilde{\Pi}_j^{\pm}(x)\big)^2 +
    \frac{1}{\tilde{K}_{\pm,j}}\big(\partial_x
    \tilde{\phi}_j^{\pm}(x)\big)^2\Big] ,
\end{equation}
and the velocities and Luttinger parameters are given by
\begin{equation}    \begin{split}
  \tilde{u}_{+,j}& =\tilde{u}_{-,j} = (u_{a,j}+u_{b,j})\sqrt{1 - \tilde{g}_{j}}  \,, \\ 
    \tilde{K}_{+,j} & = 1/\tilde{K}_{-,j} = \sqrt{\frac{1 +
      \tilde{g}_j } {1 - \tilde{g}_j}} \,,
\end{split}   \end{equation}
where
\begin{equation}
  \tilde{g}_j = \frac{g_{2;j}^{ab}(\va+\vb)}{2 \sqrt{u_{a,j} u_{b,j}/(K_{a,j}
      K_{b,j})}}
\end{equation}
The velocities are the same in the two sectors, therefore, in the most general
case only $n$ different velocities are found for the $2n$ modes. If the
couplings depend on the relative spins of the scattered electrons only, 2
different velocities are found and one recovers the usual spin-charge
separation.

\section{Strong-coupling fixed points}

In all other solutions of the scaling equations, some of the
backward-scattering, charge-transfer or umklapp processes are relevant and
scale to a strong-coupling fixed point. Although these processes may drive the
marginal forward-scattering processes, too, to the strong-coupling limit,
these fixed points will be classified by giving those backward-scattering,
charge-transfer and umklapp processes only, which become relevant.

\subsection{Fixed points}

First the backward-scattering terms are considered. It is easy to check that
the scaling equations can be satisfied if the charge-transfer and umklapp
processes are still irrelevant, they scale to zero, and only some of the
backward-scattering processes---together with some forward-scattering
ones---scale to strong coupling. FP--BS-a(b) denotes the fixed point where the
coupling of the intraband backward scattering in band $a$ (or in band $b$) has
non-vanishing value. The fixed point FP--BS-ab corresponds to the case when
the interband backward scattering is relevant and the intraband
backscatterings are irrelevant. It is possible that one or both of the
intraband and the interband backward-scattering processes are relevant at the
same time, but as we will see these fixed points do not give a new phase.

Similarly, it can be checked that the scaling equations can be satisfied by
assuming that in the fixed point, one of the charge-transfer
forward-scattering processes has non-vanishing coupling, while all
backward-scattering and umklapp processes are irrelevant. Accordingly
FP--CT-FS$\|$ and FP--CT-FS$\bot$ denote the fixed points where $g^{\rm
ct}_{\|}$ or $g^{\rm ct}_{2\bot}$, respectively are relevant, while the other
charge-transfer processes are irrelevant.

While the backward-scattering and charge-transfer forward-scattering processes
may become relevant on their own, this is not the case for the charge-transfer
backward-scattering process $\gemct$, except if $n=2$. For $n>2$, if the
charge-transfer backward-scattering process is relevant, it will necessarily
drive the interband backscattering processes ($\gemab$) and the
charge-transfer processes between electrons of parallel spin ($\gpct$), too,
to strong coupling.  The corresponding fixed point is denoted by FP--CT-BS.

Similar situation occurs when one of the bands is half-filled. In this case
the intraband umklapp processes may become relevant, and, if $n > 2$ these
processes drive the intraband backward-scattering processes, too, to strong
coupling. This fixed point is denoted by FP--U-a(b).

The names of the fixed points FP--U-abab, FP--U-abba, FP--U-ab$\|$, and
FP--CT-U, which may be reached when the system is half filled, are
self-explanatory. E.g., in the fixed point FP--U-abab, the coupling $\ghmabab$
scales to strong coupling, while in the fixed point FP--U-abba the coupling
$\ghmabba$ is relevant.  For $n=2$, these fixed points can be reached without
driving charge-transfer, backward-scattering or other umklapp processes to
strong coupling. For $n > 2$, however, necessarily some other couplings are
also relevant. E.g., the coupling $\ghmabab$ will drive the couplings $\gemab$
and $\ghp$ to strong coupling, or together with $\ghmabba$ both the intraband
backscattering and the umklapp process $\ghp$ become relevant. The
charge-transfer umklapp process drives the interband backward-scattering
processes to the strong-coupling limit.

Depending on the band filling other, more complicated fixed points may
also be reached if appropriate initial couplings are chosen. They correspond
to regions where several of the backward, charge-transfer or umklapp
processes become relevant at the same time, although they are not necessarily 
driven to strong coupling by the other relevant processes.

\subsection{The role of the individual terms}

As we have seen the two-leg fermion ladder with $n$-valued spin has two charge 
modes and $2(n-1)$ spin modes. In the strong-coupling fixed points, where one 
or more backward-scattering, charge-transfer or umklapp processes are relevant,
gap may be opened in the spectrum of some of the modes. The bosonization 
procedure allows us to determine how many of the modes are gapped and how many 
remain gapless. For this we reexpress the one-particle fermion fields and the 
relevant scattering processes in terms of the boson degrees of freedom. 

The continuum fermion fields of the right and left moving particles are
defined by 
\begin{equation}  \begin{split}
a_{+,k,\sigma}& = \frac{1}{\sqrt{L}}\int \psi_{+,\sigma}^a(x) e^{-i(\ka+k)x} 
\d x, \\
a_{-,k,\sigma}& = \frac{1}{\sqrt{L}}\int \psi_{-,\sigma}^a(x) e^{-i(-\ka+k)x} 
\d x,
\end{split}
\end{equation}
and by similar formulas for the operators $b$ which annihilate left or
right moving particles in band $b$. The bosonic phase fields are related to the
fermionic fields by the relations
\begin{equation}
\begin{split}
    \psi_{+,\sigma}^{a(b)}(x) & = \frac{1}{2\pi} F_{+,\sigma}^{a(b)}
        :e^{i\sqrt{4\pi} \phi_{+,\sigma}^{a(b)}(x)}:\,,\\ 
    \psi_{-,\sigma}^{a(b)}(x) & = \frac{1}{2\pi} F_{-,\sigma}^{a(b)} 
        :e^{i\sqrt{4\pi} \phi_{-,\sigma}^{a(b)}(x)}:\,.
\end{split}
\end{equation}
Here $F_{\pm,\sigma}^{a(b)}$ are the Klein factors which make sure that the
fermion operators satisfy the proper anticommutation relations even though the
phase fields $\phi_{\pm,\sigma}^{a(b)}$ satisfy bosonic commutation relations
and they commute with the Klein factors. The total bosonic fields and their
duals are:
\begin{equation}
  \begin{split}
    \phi_\sigma^{a(b)} & = \phi_{+,\sigma}^{a(b)} + \phi_{-,\sigma}^{a(b)}, \\
    \theta_\sigma^{a(b)} & = \phi_{+,\sigma}^{a(b)} - \phi_{-,\sigma}^{a(b)}.
  \end{split}
\end{equation}
These fields are related to the charge and spin fields defined earlier
by the relations
\begin{equation}   \begin{split}
     \phi_{\rm c}^{a(b)} & = \frac{1}{\sqrt{n}}\sum_{\sigma=1}^n
     \phi_{\sigma}^{a(b)},\\ \phi_{m{\rm s}}^{a(b)} & =
     \frac{1}{\sqrt{m(m+1)}}\left(\sum_{\sigma=1}^m \phi_{\sigma}^{a(b)} -
     m\phi_{m+1}^{a(b)}\right),
\end{split}   \end{equation}
with $m=1,\dots,n-1$.

In order to see what the role of those scattering processes is which should
vanish in the Tomonaga-Luttinger fixed point and which may drive the system
into a strong-coupling fixed point, we rewrite them in this continuum boson
representation. The true backward-scattering processes are proportional to
\begin{equation}  \begin{split}
  & \gema \int \bigg[ \sum_{\sigma\neq \sigma'} \cos \big(\sqrt{4\pi}
    (\phi_\sigma^a - \phi_{\sigma'}^a)\big)\bigg] \d x\, ,\\ & \gemb \int
    \bigg[ \sum_{\sigma\neq \sigma'} \cos \big(\sqrt{4\pi} (\phi_\sigma^b -
    \phi_{\sigma'}^b)\big)\bigg] \d x\, ,\\ & \gemab \int
    \bigg[\sum_{\sigma\neq\sigma'} \cos \big(\sqrt{\pi} (\phi_{\sigma}^a -
    \phi_{\sigma'}^a + \phi_{\sigma}^b -\phi_{\sigma'}^b)\big) \\ & \hskip
    1.2cm \times \cos\big( \sqrt{\pi} (\theta_\sigma^a -\theta_{\sigma'}^a
    -\theta_\sigma^b +\theta_{\sigma'}^b)\big)\bigg] \d x \, .
\end{split}   \end{equation}
It follows from the sine-Gordon form of the interactions that they may give
rise to gapped soliton solutions either in the charge or spin
sector. It is easy to see that the backward-scattering terms contain spin
modes only. For this one has to realize that the combinations 
$\phi_\sigma^{a(b)}-\phi_{\sigma'}^{a(b)}$ (and the similar combinations of 
the dual fields) occurring in the cosine functions are orthogonal to the
charge modes $\phi_{\rm c}^{a(b)}$ ($\theta_{\rm c}^{a(b)}$). The intraband 
backward-scattering process opens $n-1$ spin gaps while the interband process 
($\gemab$) opens $2(n-1)$ spin gaps.  

In boson representation, except for a factor, the charge-transfer processes
have the form
\begin{equation}  \begin{split} 
   & \gpct \int \bigg[ \sum_{\sigma} \cos \big(\sqrt{4 \pi}(\theta_\sigma^a -
    \theta_\sigma^b)\big)\bigg]\d x\, ,\\ & \gkmct \int\bigg[ \sum_{\sigma\neq
    \sigma'} \cos \big(\sqrt{\pi} (\theta_\sigma^a +\theta_{\sigma'}^a -
    \theta_\sigma^b - \theta_{\sigma'}^b) \big) \\ & \phantom{+++} \times \cos
    \big(\sqrt{\pi} (\phi_\sigma^a -\phi_{\sigma'}^a - \phi_\sigma^b +
    \phi_{\sigma'}^b) \big) \bigg] \d x\, , \\ & \gemct \int\bigg[
    \sum_{\sigma\neq \sigma'} \cos \big(\sqrt{\pi} (\theta_\sigma^a
    +\theta_{\sigma'}^a - \theta_\sigma^b - \theta_{\sigma'}^b) \big) \\ &
    \phantom{+++} \times \cos \big(\sqrt{\pi} (\phi_\sigma^a -\phi_{\sigma'}^a
    + \phi_\sigma^b - \phi_{\sigma'}^b) \big) \bigg]\d x\, .
\end{split}  \end{equation}
The first two expressions which describe charge-transfer
forward-scattering processes, contain one charge mode and $n-1$ spin modes,
moreover, they couple the same modes, and therefore they modify the
spectrum in the same way.  For this reason we will not distinguish between
these processes and the notation FP--CT-FS will be used for any of the two
strong-coupling fixed points FS--CT-FS$\|$ and FS--CT-FS$\bot$ or for the
fixed hypersurface FP--[CT-FS$\parallel$]+[CT-FS$\perp$] on which both $\gpct$
and $\gkmct$ are relevant. In the expression for the charge-transfer
backward-processes the first cosine function contains one charge mode and
$n-1$ spin modes if $n > 2$, while if $n=2$ it contains only a charge
mode. The second cosine function contains the other $n-1$ spin modes. As a
result one charge mode and one spin mode appears in the $\gemct$ term if $n=2$,
but one charge and $2(n-1)$ spin modes for $n>2$. 

The umklapp terms may also be responsible for the opening of a gap in the 
excitation spectrum.  The intraband umklapp processes have the form:
\begin{equation}   \begin{split}
  & \ghma \int \bigg[ \sum_{\sigma\neq \sigma'} \cos \big(\sqrt{4\pi}
  (\phi_\sigma^a + \phi_{\sigma'}^a)\big)\bigg]\d x\, ,\\
  & \ghmb \int \bigg[ \sum_{\sigma\neq \sigma'} \cos \big(\sqrt{4\pi}
  (\phi_\sigma^b + \phi_{\sigma'}^b)\big)\bigg]\d x\, .
\end{split}    \end{equation}
These umklapp processes do not contain spin modes but one charge modes only if
$n=2$, while if $n > 2$ the charge mode is coupled to $n-1$ spin modes. 

The interband umklapp processes take the form
\begin{equation}   \begin{split}
   & \ghmabab \int \bigg[ \sum_{\sigma\neq \sigma'} \cos \big(\sqrt{\pi}
  (\theta_\sigma^a -\theta_{\sigma'}^a - \theta_\sigma^b + \theta_{\sigma'}^b)
  \big)\\ 
     & \hskip 1.2cm \times \cos \big(\sqrt{\pi}
  (\phi_\sigma^a +\phi_{\sigma'}^a + \phi_\sigma^b + \phi_{\sigma'}^b)
  \big)  \bigg] \d x\, ,\\
  & \ghmabba \int \bigg[ \sum_{\sigma\neq \sigma'} \cos \big(\sqrt{4\pi}
  (\phi_\sigma^a + \phi_{\sigma'}^b)\big)\bigg] \d x\, ,\\
  & \ghp \int \bigg[\sum_{\sigma} \cos \big(\sqrt{4\pi}
  (\phi_\sigma^a+ \phi_\sigma^b)\big)\bigg] \d x\, , \\
  &\ghmct\int \bigg[\sum_{\sigma\neq\sigma'} \cos \big(\sqrt{\pi}
  (\phi_{\sigma}^a+\phi_{\sigma'}^a + \phi_{\sigma}^b+\phi_{\sigma'}^b) \\ 
  & \hskip 1.2cm \times \cos \big(\sqrt{\pi}(\theta_{\sigma}^a + 
  \theta_{\sigma'}^a - 
  \theta_{\sigma}^b-\theta_{\sigma'}^b)\big)\bigg]\d x \, .
\end{split}    \end{equation}
The interband umklapp scattering $\ghp$ involves one charge and $n-1$ spin
modes for any $n$. For the other terms different behavior is found for $n=2$
and $n > 2$. The processes $\ghmabab$ and $\ghmabba$ involve the same modes,
namely they open one charge gap and $2(n-1)$ spin gaps for $n > 2$, however,
only one charge gap and one spin gap are opened if $n=2$. Therefore similarly
to the charge-transfer forward scatterings they (alone or together)
produce the same gap structure in the spectrum. In what follows these
processes will not be distinguished and the notation FP--U-ab$\perp$ will be
used for any of the fixed points FP--U-abab and FP--U-abba or for the
hypersurface on which $\ghmabab$ and $\ghmabba$ are relevant simultaneously
(FP--[U-abab]+[U-abba]).

The term corresponding to charge-transfer
umklapp processes contains only two charge modes if $n=2$, but two charge and
2$(n-1)$ spin modes when $n>2$, i.e., all modes are gapped. 

\subsection{Possible phases}

Following Balents and Fisher \cite{LBMPAF-2} the phases belonging to the
various fixed points will be characterized by the number of gapless charge and
spin modes, respectively. 

In the attractive regime of the Tomonaga-Luttinger fixed point where all
backward-scattering, charge-transfer, and umklapp processes are irrelevant, 
their couplings scale to zero, both charge modes and all spin modes are 
gapless. In this C2S$2(n-1)$ phase the system is a $2n$-component
Luttinger liquid.

As it has been pointed out earlier, the backward-scattering processes are 
responsible for opening gaps in the spin sector, while the charge-transfer and 
umklapp processes may open gaps both in the charge and spin sectors. Because 
of the special behavior for $n=2$ this case will be discussed separately. 

For generic filling and for $n>2$ the charge sector remains gapless if the
charge-transfer processes are irrelevant. Depending on whether one of the
intraband or the interband backward-scattering processes is relevant, and the
fixed point FS--BS-a(b) or FS--BS-ab is reached, a phase C2S$(n-1)$ or C2S0 is
found, as shown in Table \ref{tab:phases1}. Since all spin modes are gapped if
the interband backward-scattering is relevant, the gap structure is not
modified by the intraband backward-scattering processes. The same phase is
obtained irrespective whether these processes are relevant or not. Similarly,
because the two intraband backward-scattering terms open gaps in different
spin modes, the spin sector is fully gapped when both of them are
relevant. The phase is not modified by a relevant or irrelevant interband
backward scattering.

\begin{table}[htb]
  \centering
  \begin{tabular}{|c|c|} \hline
     fixed point & phase \\ \hline \hline
     FP--TL & \,\, C2S$2(n-1)$ \,\, \\ \hline  
      FP--BS-a(b) & C2S$(n-1)$ \\ \hline 
     FP--[BS-a]+[BS-b] & {\raisebox{-2mm}[0mm][0mm]{C2S0}} \\
      FP--BS-ab &  \\  \hline
      FP--CT-FS & C1S$(n-1)$ \\  \hline
      FP--CT-BS  & {\raisebox{-2mm}[0mm][0mm]{C1S0}} \\ 
    \,\, FP--[CT-FS]+[BS-a(b,ab)] \,\, &   \\  \hline
  \end{tabular}
 \caption{Possible fixed points and phases of the $n$-component fermion ladder 
for $n > 2$ at generic filling.}
  \label{tab:phases1}
\end{table}

The phase realised in the regime where the charge-transfer forward-scattering
processes are relevant, can be described as C1S$(n-1)$. Finally if both the
backward-scattering and charge-transfer processes are relevant, the spin
sector is fully gapped and only one charge mode is gapless (C1S0). Due to the
fully gapped spin sector in the fixed point FP--CT-BS ($\gemct$ is relevant),
the same phase is obtained for $n > 2$ even if the other backward-scattering
processes become relevant.

As we see there is always at least one gapless charge mode, and a gapped
charge mode necessarily implies that some of the spin modes become gapped.

The situation is similar for $n=2$, except that if the charge-transfer
backward-scattering processes are relevant, they lead to phase
C1S1. However, if either the charge-transfer backward and forward-scattering
processes are relevant simultaneously (in this case the couplings related to
other backward-scattering processes may be relevant or irrelevant) or the
charge-transfer backward-scatterings and at least one of the other
backward-scatterings are relevant simultaneously, the whole spin sector is
gapped. The phases corresponding to the various fixed points are given in
Table \ref{tab:phases1n=2}.

\begin{table}[htb]
  \centering
  \begin{tabular}{|c|c|} \hline 
       fixed point & phase \\  \hline \hline
     FP--TL & \,\, C2S2 \,\, \\  \hline
     FP--BS-a(b) & C2S1 \\ \hline 
    FP--[BS-a]+[BS-b] & {\raisebox{-2mm}[0mm][0mm]{C2S0}} \\
     FP--BS-ab & \\ \hline 
    FP--CT-BS & {\raisebox{-2mm}[0mm][0mm]{C1S1}} \\
     FP--CT-FS &  \\ \hline
    \,\, FP--[CT-BS]+[BS-a(b,ab)] \,\, &   
    \\ \,\, FP--[CT-BS]+[CT-FS] 
    \,\, & C1S0  \\
    \,\, FP--[CT-FS]+[BS-a(b,ab)] \,\, & \\    \hline
  \end{tabular}
  \caption{Possible fixed points and phases of the $n$-component fermion ladder 
    for $n=2$ at generic filling.}
  \label{tab:phases1n=2}
\end{table}

When one of the subbands is half filled, and the corresponding intraband 
umklapp processes are relevant, the coupling $\ghma$ (or $\ghmb$) 
opens one charge gap and $n-1$ spin gaps. If the backward-scattering
processes become relevant simultaneously, they open gaps in the remaining
gapless spin modes, while charge-transfer together with the umklapp
processes opens gaps in all branches of the spectrum. This, of course, is not
modified if besides the umklapp and charge-transfer processes other couplings,
e.g., backward scatterings become relevant. The possible phases 
are given in Table \ref{tab:phases2}.

\begin{table}[htb]
  \centering
  \begin{tabular}{|c|c|} \hline 
         fixed point & phase \\  \hline \hline
     FP--TL & \,\, C2S$2(n-1)$ \,\, \\ \hline 
     FP--U-a & C1S$(n-1)$ \\ \hline 
    FP--[U-a]+[BS-b(ab)] & C1S0 \\  \hline 
    \,\, FP--[U-a]+[CT-BS]  \,\, & {\raisebox{-2mm}[0mm][0mm]{C0S0}} \\
    \,\, FP--[U-a]+[CT-FS] \,\,  &  \\ \hline
  \end{tabular}
  \caption{Possible fixed points and phases of the $n$-component fermion ladder
    with half-filled subband for $n > 2$.}
  \label{tab:phases2}
\end{table}

The phases obtained for $n=2$ are given in Table \ref{tab:phases2n=2}.  The
difference compared to $n>2$ is due to the fact that the intraband
umklapp processes open gap in the charge sector only and therefore phases with
a fully gapless spin sector and one soft charge mode (C1S2) and a fully gapped
charge sector with one soft spin mode (C0S1) are also possible. Similarly to
the $n>2$ case only one charge gap can be found in the spectrum if the 
charge-transfer processes are irrelevant (phases C1S1 and C1S0).

\begin{table}[htb]
  \centering
  \begin{tabular}{|c|c|} \hline 
         fixed point & phase \\  \hline \hline
    FP--TL & \,\, C2S2 \,\,\\ \hline 
      FP--U-a & C1S2 \\ \hline
    FP--[U-a]+[BS-a(b)] & C1S1 \\ \hline \,\,
    FP--[U-a]+[BS-a]+[BS-b] \,\, & {\raisebox{-2mm}[0mm][0mm]{C1S0}} \\ \,\,
    FP--[U-a]+[BS-ab] \,\, &  \\ \hline
    FP--[U-a]+[CT-BS] & {\raisebox{-2mm}[0mm][0mm]{C0S1}} \\ 
     FP--[U-a]+[CT-FS] & \\ \hline
    \,\, FP--[U-a]+[CT-BS]+[BS-a(b,ab)] \,\, & \\
    \,\, FP--[U-a]+[CT-BS]+[CT-FS] \,\, & C0S0 \\ \,\,
    FP--[U-a]+[CT-FS]+[BS-a(b,ab)] \,\, & \\ \hline
  \end{tabular}
  \caption{Possible fixed points and phases of the $n$-component fermion ladder
    with half-filled subband for $n=2$.}
  \label{tab:phases2n=2}
\end{table}

The phases occurring when the system is half-filled and one or more interband
umklapp processes are relevant are listed in Table \ref{tab:phases3} for $n >
2$. The umklapp scattering processes---similarly to the charge-transfer
processes---open at least one charge and $n-1$ spin gaps. A soft charge mode
may be present in the spectrum only if all charge-transfer processes are
irrelevant, moreover, in phase C1S$(n-1)$ the umklapp scattering between
electrons with parallel spins only may be relevant. As it was mentioned in the
previous section the couplings $\ghmabab$ and $\ghmabba$ open $2(n-1)$ spin
gaps, therefore---due to the fully gapped spin sector---the relevance of the
processes which open gap in the spin sector only ($\gema$, $\gemb$, $\gemab$)
does not modify the spectrum. However, in phase C1S0, besides the coupling
$\ghp$ at least one backward scattering has to be scaled to strong
coupling. We note that the charge-transfer umklapp process $\ghmct$ alone
makes the spectrum fully gapped, but of course in phase C0S0, all couplings
may be relevant.

\begin{table}[htb]
  \centering
  \begin{tabular}{|c|c|}  \hline 
        fixed point & phase \\  \hline \hline
    FP--TL & \,\, C2S$2(n-1)$ \,\, \\ \hline 
      FP--U-ab$\parallel$ & C1S$(n-1)$ \\ \hline 
      FP--U-ab$\perp$ & {\raisebox{-2mm}[0mm][0mm]{C1S0}} \\ \,
    FP--[U-ab$\parallel$]+[BS-a(b,ab)] \, & \\ \hline FP--CT-U & \\ \,
    FP--[U-ab$\perp$]+[CT-FS] \, & \\ \, FP--[U-ab$\perp$]+[CT-BS] \, & C0S0
    \\ \, FP--[U-ab$\|$]+[CT-FS] \, & \\ \, FP--[U-ab$\|$]+[CT-BS] \, & \\
    \hline
  \end{tabular}
  \caption{Possible fixed points and phases of the half-filled 
    $n$-component fermion ladder for $n > 2$.}
  \label{tab:phases3}
\end{table}

If $n=2$ the charge-transfer umklapp processes do not open gap in the spin
sector, therefore, new phases---namely C0S2 and C0S1---occur in the phase
diagram compared to $n > 2$ (see Table \ref{tab:phases3n=2}). It
has to be noted that similarly to $n > 2$ the spin sector can be
fully gapped without relevant backward-scattering processes 
(FP--[U-ab$\perp$]+[U-ab$\|$]) and the relevance or irrelevance
of the non-charge-transfer backward scattering processes does not modify the
gap structure. Moreover, in the fixed points, where relevant
charge-transfer umklapp processes are present, the relevance of any other
coupling leads to phase C0S1 (the last three fixed points for phase
C0S1), except for $\gemab$ which opens $2(n-1$) spin gaps, so the
spectrum becomes fully gapped and the corresponding phase is C0S0.

\begin{table}[htb]
  \centering
  \begin{tabular}{|c|c|} \hline 
          fixed point & phase \\  \hline \hline
     FP--TL & \,\, C2S2 \,\, \\ \hline 
      FP--U-ab$\perp$(ab$\|$)  & C1S1 \\  \hline 
   \,\, FP--[U-ab$\perp$]+[BS-a(b,ab)]   \,\, & \\ 
\,\, FP--[U-ab$\perp$]+[U-ab$\|$]  \,\, &   C1S0 \\
\,\, FP--[U-ab$\|$]+[BS-a(b,ab)] \,\, & \\  \hline 
     FP--CT-U & C0S2 \\  \hline 
\, FP--[U-ab$\perp$]+[CT-FS] \,  & \\ 
\, FP--[U-ab$\|$]+[CT-BS] \,  & \\ 
   \,\, FP--[CT-U]+[BS-a(b,CT)]  \,\, & C0S1 \\ 
    \,\, FP--[CT-U]+[U-ab$\perp$(ab$\|$)]  \,\, &  \\
    \,\, FP--[CT-U]+[CT-FS] \,\, & \\  \hline 
 \,\, e.g. FP--[U-ab$\perp$]+[CT-BS]  \,\, & C0S0  \\
    \hline
  \end{tabular}
  \caption{Possible fixed points and phases of the half-filled 
    $n$-component fermion ladder for $n=2$.}
  \label{tab:phases3n=2}
\end{table}

It can be seen that all possible phases have been found if $n=2$ while for
$n>2$ some theoretically possible phases do not occur. This is due to the fact
mentioned above that in this case there are no couplings which open gap in the
charge sector only:\ a charge gap opens always together with spin gaps. A more
complete phase diagram could be obtained if the couplings have a more general
spin dependence. Although, even in this case the charge gaps are coupled to
spin gaps, it would be possible to open a single spin gap---if the
corresponding fixed point exists---contrary to case investigated above where
$n-1$ spin gaps occur always together.

\section{Numerical results}

The behavior close to the weak-coupling fixed point has been
analyzed in Sec.\ IV analytically. A similar calculation close to the
strong-coupling fixed points is not possible since---as mentioned
earlier---the scaling equations derived in the leading logarithmic
approximation are not valid in the strong-coupling regime. The relevance or
irrelevance of the couplings could, however, be deduced from the numerical
solution of the scaling equations. We have done numerical studies for
different band fillings for models in which the bare couplings correspond to a
repulsive Hubbard ladder, and also for somewhat more general situations
characterized by several bare coupling constants. 

\subsection{Hubbard ladder}
\label{sec:Hubb}

In a Hubbard ladder the initial couplings of all intraband and interband
scattering processes $g_{i\perp}^{\alpha\beta\gamma\delta}$ are related to the
same on-site interaction $U$ defined in Eq.\ (\ref{eq:H-chain}) while
$g_{\parallel}^{\alpha\beta\gamma\delta}$ and $\ghp$ are equal to zero.  The
numerical results presented below were obtained for $U/(2 \pi \hbar v_a) =
0.1$ and different values for the ratio $v_b/v_a$. Other choices for $U/(2 \pi
\hbar v_a)$ gave qualitatively identical results.

\subsubsection{Generic filling}

\begin{figure}[htb]
  \centering
  \includegraphics[scale=0.35]{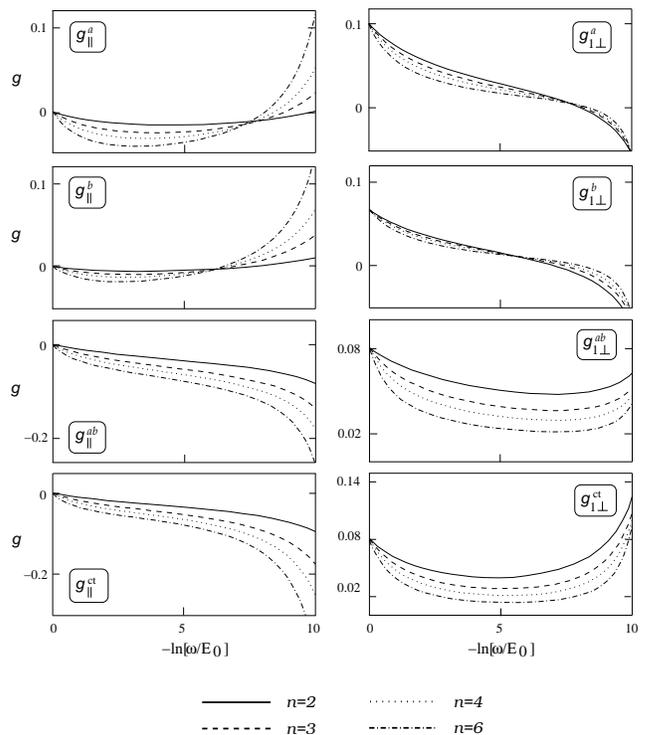}
  \caption{The second-order scaling curves for a weak-coupling Hubbard ladder
     ($U/(2\pi\hbar v_a)=0.1$, $\vb/\va=1.5$) for the 8 independent couplings
     and for different values of $n$.}
  \label{fig:H1}
\end{figure}

When the umklapp processes may be neglected 12 scattering processes have to be
taken into account. Looking at the lowest-order scaling equations it can be
shown that if the initial couplings statisfy the relations
\begin{equation}
\label{eq:inv-gf}
  g_{1\perp}^{a(b,ab,{\rm ct})}=g_{\parallel}^{a(b,ab,{\rm ct})}
  + g_{2\perp}^{a(b,ab,{\rm ct})},
\end{equation}
what is the case for the Hubbard ladder, the same relations remain valid for
the renormalized couplings. Due to these scale-invariant surfaces eight
independent couplings describe the system. The second-order scaling curves
for these couplings for $v_b/v_a = 1.5$ are shown in Fig.\ \ref{fig:H1}.

One can see that all scaling trajectories go to infinity so all couplings are
relevant for all value of $n$. The weak-coupling fixed point, which is stable
for the $n=2$ Hubbard chain, is not reached for the Hubbard ladder. Since the
backward-scattering and charge-transfer processes are relevant the
corresponding phase is C1S0. For generic filling the SU($n$) Hubbard ladder is
not a Luttinger liquid.

Similar results are obtained for other $\gamma$ values provided they are of
the order of unity. Solving the scaling equations formally for larger values
of $\gamma$, it was found that successively some of the charge-transfer 
and backward-scattering processes become irrelevant and new phases occur. 
Namely, if $\vb/\va$ or $\va/\vb$ (since $\gamma$ takes the same value when 
$\vb/\va$ is replaced by $\va/\vb$) is approximately 8.3, all charge-transfer 
processes, the interband backward scattering $g_{1\perp}^{ab}$ and one 
of the intraband backward-scattering terms $g_{1\perp}^{a}$ or 
$g_{1\perp}^{b}$, belonging to the band with smaller velocity, become 
irrelevant and FP--BS-b or FP--BS-a is reached. The corresponding phase 
is C2S($n-1$). The surviving intraband backscattering $g_{1\perp}^{b}$ or 
$g_{1\perp}^{a}$ becomes irrelevant above an even higher 
value---approximately $8.7$---of the velocity ratio $\vb/\va$ or $\va/\vb$. 
None of the couplings that could open gap in the spectrum seem to be relevant 
and the Luttinger liquid phase C2S2$(n-1)$ would be obtained. These boundaries 
vary little with $n$. These phases appear, however, in such situations where one 
of the subbands is almost empty or almost full, the curvature of the dispersion 
relation may be important, and a model with linearized spectrum may not be a 
good approximation. For this reason these phases are not indicated in the phase diagram.

\subsubsection{Role of intraband umklapp processes}

\begin{figure}[htb]
  \centering
  \includegraphics[scale=0.35]{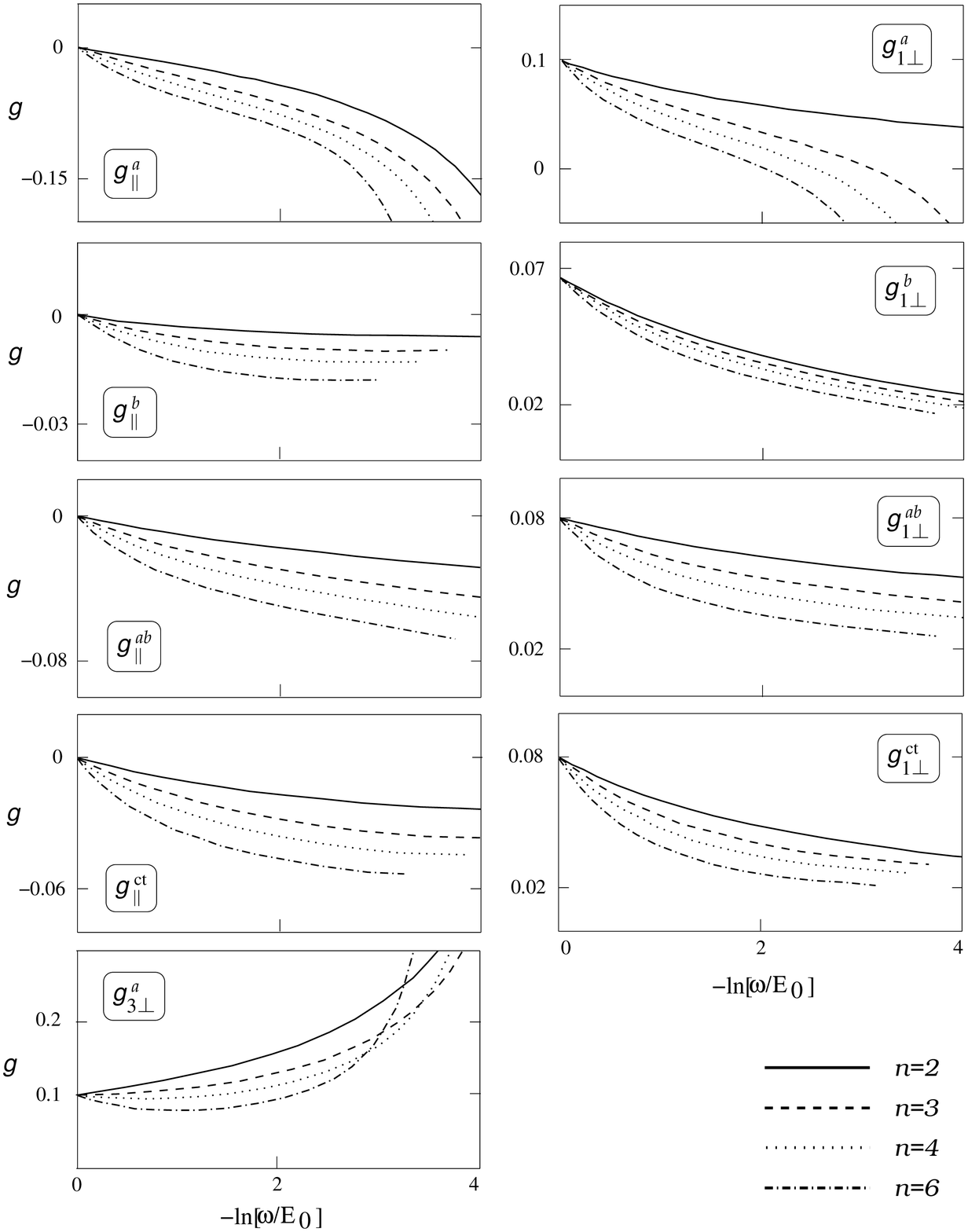}
  \caption{The second-order scaling curves for the Hubbard ladder 
    ($U/(2\pi\hbar v_a)=0.1$, $\vb/\va=1.5$) with half-filled subband.}
  \label{fig:H2}
\end{figure}

If $4\ka$ is equal to a reciprocal lattice vector $G$ and the umklapp process
$\ghma$ is relevant the surfaces determined by \eqref{eq:inv-gf} remain
invariant. Therefore out of the 13 couplings which give contribution 9 are 
independent. The scaling curves for these couplings obtained in the leading
logarithmic approximation can be seen in Fig.\ \ref{fig:H2}.

It is found that starting from the bare couplings characterizing the Hubbard
ladder the weak-coupling fixed point cannot be reached for any value of $n$
and for any value of $\gamma$. The strong-coupling fixed point
FP--[U-a]+[CT-FS] where the umklapp processes are relevant together with the
charge-transfer ones but all backward-scattering processes are irrelevant
seems to be accessible for the Hubbard ladder if $n=2$ (independently of the
value of $\gamma$). The corresponding phase is C0S1. As has been seen above,
this fixed point could not be reached for $n > 2$ and phase C0S0 is realized
there.  Similar result has been found\cite{ESZJS} for the 1D
SU($n$) chain where it was shown that the umklapp scattering suppresses the
normal backward-scattering processes for $n=2$ only.

\subsubsection{Half-filled system: interband umklapp processes}

\begin{figure}[htb]
  \centering
  \includegraphics[scale=0.35]{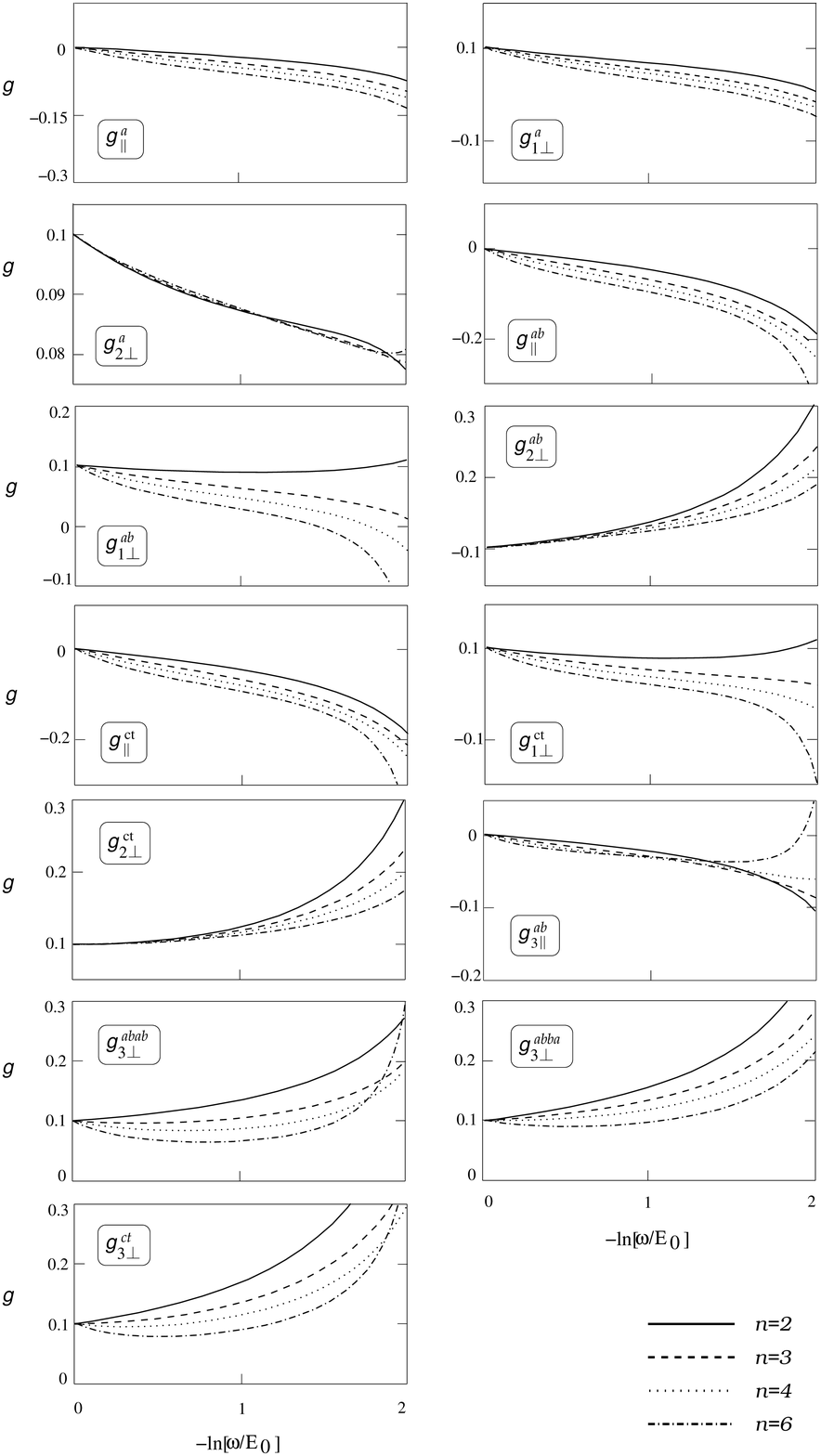}
  \caption{The second-order scaling curves for the half-filled Hubbard ladder
($U/(2\pi\hbar v_a)=0.1$).}
  \label{fig:H3}
\end{figure}

When the whole system is half-filled, the contribution of the interband
umklapp processes should be taken into account. The scaling equations are
invariant under the interchange of bands $a$ and $b$
\begin{equation}
g_i^a=g_i^b
\end{equation}
and therefore out of the 16 coupling which give
contribution in a half-filled system, 13 are independent. The scaling curves
for these couplings are shown in Fig.\ \ref{fig:H3}.

The weak-coupling fixed point determined by (\ref{eq:fixp1}),
(\ref{eq:fixp1kieg}) and (\ref{eq:fixp3}) cannot be reached starting from the
Hubbard ladder. Similarly to the case of generic filling, for any $n$ and
$\gamma$ all couplings go to infinity. But since now the umklapp processes,
too, are relevant, the corresponding phase is C0S0.

\begin{figure}[htb]
  \centering
  \includegraphics[scale=0.3]{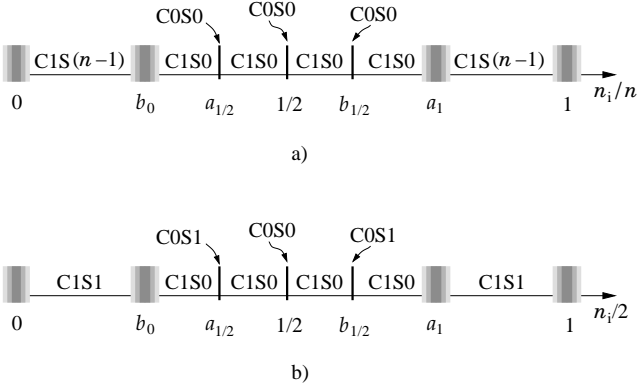}
  \caption{The phase diagram of the Hubbard ladder: a) for $n>2$ and b) for
  $n=2$.}
  \label{fig:H4}
\end{figure}

Figure \ref{fig:H4} shows the phase diagram of the Hubbard model. The control
parameter is the filling $n_i/n$ where $n_i$ is the particle number per
site. $n_i/n=0$ and 1 denote the totally empty and full system, respectively,
$b_0$ denotes the filling when the upper band is almost empty, similarly $a_1$
denotes the filling when the lower band is almost full. Close to these
fillings the linearization of the spectrum is not valid, and the shaded regime
around these points indicates that the results presented in this paper using a
linearized spectrum may not be reliable at these fillings. For this reason 
the phases found in these regimes are not indicated in the figure.

The upper figure (a) shows the phase diagram of the $n$-component Hubbard
ladder for $n>2$. Figure (b) is valid for 2-component fermions. As one can see 
the $n=2$ and $n>2$ models behave differently in the presence of intraband
umklapp processes. 

\subsection{Chains with forward scattering only} 

As another special case we consider two chains with nonchiral intrachain
forward scattering only. Although the interband scattering processes that
arise when the one-particle transverse hopping is switched on are all of
forward-scattering type, the charge-transfer processes do not conserve 
the chiral charge and spin. For this reason in general, such a ladder does 
not behave like a Luttinger liquid.

The couplings which play a role in this coupled system are related to
scattering processes between electrons of the same spin: $\gpa$, $\gpb$,
$\gpab$ and $\gpct$, and to forward scattering processes between electrons of
different spins: $\gkma$, $\gkmb$, $\gkmab$ and $\gkmct$. The scaling
equations for the intraband scatterings are:
\begin{subequations}
\begin{flalign}
  \frac{\d \gpa}{\d \ln x} =  & \, - \gamma\gpctn  ,\\
  \frac{\d \gkma}{\d \ln x} = & \,   \gamma\gkmctn 
\end{flalign}
\end{subequations}
Similar equations hold for the couplings in band $b$. The Lie equations for
the interband scatterings without charge transfer are:
\begin{subequations}
\begin{flalign}
  \frac{\d \gpab}{\d \ln x} = & \, \gpctn ,\\ 
  \frac{\d \gkmab}{\d \ln x} = & \, - \gkmctn ,
\end{flalign}
\end{subequations}
and for the charge-transfer processes
\begin{subequations}
\begin{flalign}
 \frac{\d \gpct}{\d \ln x} = & \,- \gpct\Big[\gpa + \gpb - 2\gpab\Big], \\ 
 \frac{\d \gkmct}{\d \ln x} = & \, \gkmct\Big[\gkma + \gkmb - 2\gkmab\Big] .
\end{flalign}
\end{subequations}
As can be seen, the coupling constants of the scattering processes between
electrons of the same spin and of different spins are not coupled, the two
scattering channels are separated, and analogous equations are obtained.

These equations have one weak-coupling fixed point only, namely when the
processes which transfer charge from one band to the other are
irrelevant. This fixed ``point'' is a six-dimensional hypersurface in the
parameter space and it can be attractive if
\begin{subequations}
\label{eq:req_cTL1}
\begin{flalign}  
  2\gpabast \geq & \, \gpaast + \gpbast, \\
  \gkmaast + \gkmbast \geq & \, 2\gkmabast.
\end{flalign}
\end{subequations}
As we have seen in Sec.\ \ref{sec:fixp}, at this fixed point the
system is a Luttinger liquid. One finds, however, different behavior depending
on whether the interband scatterings are scaled out of the problem or not. If
these processes vanish, the two charge modes become independent, and similarly
the spin modes are decoupled, while if the interchain couplings survive, the
charge modes become coupled and have identical velocities. Similar situation
occurs for the spin modes as well.

The scaling equations have been solved numerically for one-parameter models, 
i.e., for models characterized by a single bare coupling constant $U$. According
to \eqref{eq:coupl-dimless} the dimensionless couplings 
$g^{\alpha\beta\gamma\delta}$ appearing in the scaling equations are
$\Tilde{U}=2U/\pi\hbar(v_\alpha+v_\beta+v_\gamma+v_\delta)$. 

We have found, starting from a one-parameter Hubbard-like model---the 
Coulomb interaction between electrons of the same spin is zero---that the 
charge-transfer processes become relevant, the stable fixed point is 
FP--CT-FS both for fully repulsive and fully attractive models.
In fact the sign of the couplings $\gpct$ and $\gkmct$ is not relevant in this
respect, so this strong-coupling fixed point is reached if
\begin{equation}
       g_{\|}^{a(0)} =  g_{\|}^{b(0)} =  g_{\|}^{ab(0)} =  g_{\|}^{{\rm ct}(0)} =  0,
\end{equation}
and 
\begin{equation}
    g_{2\bot}^{a(0)} = g_{2\bot}^{b(0)} = g_{2\bot}^{ab(0)} = \pm 
      g_{2\bot}^{{\rm ct}(0)} = \pm \Tilde{U}. 
\end{equation}
One charge and $n-1$ spin gaps open in the spectrum, the phase is C1S(n-1). 

Due to the analogous structure of the Lie-equations listed above the same 
behavior is found if the bare couplings $g_{\|}^{(0)}$ are all
repulsive or attractive while the bare $g_{2\bot}^{(0)}$ couplings vanish:
\begin{equation}
       g_{\|}^{a(0)} =  g_{\|}^{b(0)} =  g_{\|}^{ab(0)} =  \pm 
         g_{\|}^{{\rm ct}(0)}= \pm \Tilde{U}
\end{equation}
and
\begin{equation}
    g_{2\bot}^{a(0)} = g_{2\bot}^{b(0)} = g_{2\bot}^{ab(0)} = \pm 
      g_{2\bot}^{{\rm ct}(0)} = 0. 
\end{equation}
Moreover, if the couplings $g_{\|}^{(0)}$ or $g_{2\bot}^{(0)}$ are all
repulsive or attractive, the gap structure is not modified when some of
the other bare couplings appear with opposite sign. 

Luttinger liquid behavior could, however, be found when the intraband 
forward scatterings ($\gkma$ and $\gkmb$) are repulsive and the interband 
forward scattering ($\gkmab$) is attractive, 
\begin{equation}
       g_{\|}^{(0)} = 0 , \,\,\,\,  g_{2\bot}^{a(0)} = g_{2\bot}^{b(0)}=
    - g_{2\bot}^{ab(0)}= \pm g_{2\bot}^{{\rm ct}(0)} = \Tilde{U},
\end{equation}
since in this case the charge-transfer processes scale to zero. As we have 
seen in Sec.\ IV two different velocities appear in the system. Similar 
situation is encountered if the intraband forward scattering between particles 
of the same spin is attractive and the intraband forward scattering is repulsive,
\begin{equation}
       g_{\bot}^{(0)} = 0 , \,\,  g_{2\|}^{a(0)} = g_{2\|}^{b(0)}=
    - g_{2\|}^{ab(0)}= \pm g_{2\bot}^{{\rm ct}(0)} = -\Tilde{U}.
\end{equation}

These results are independent of the value of $n$. Although
the Luttinger-liquid behavior with four different velocities is theoretically
possible, for the parameter values chosen no such behavior has been found, 
because even though the charge-transfer processes are scaled out, the
interband forward-scattering processes $g_\|^{ab}$ and $g_{2\perp}^{ab}$ 
remain finite.

\section{Conclusions}
\label{sec:summ}

In the present paper, we have investigated a fermion ladder in which two
$n$-component fermion chains are coupled by interchain hopping. Using the
multiplicative renormalization-group procedure we have derived the second 
order scaling equations, determined the fixed points and analyzed the
possible phases. 

One weak-coupling fixed point has been found where the system behaves like a
Luttinger liquid. In this fixed point, the processes which violate the
conservation of the particle number and spin on each branch separately, namely
the charge-transfer processes from one band to the other, the backward and the
umklapp processes are irrelevant. In the basin of attraction of the
Tomonaga-Luttinger fixed point, where all modes, namely the two charge and
2($n-1$) spin modes are soft, the Hamiltonian has been diagonalized for
general spin-dependence of the couplings.  In this fixed point, different
behavior has been found depending on the relevance or irrelevance of the
interband forward scatterings. If they are irrelevant the system behaves like
two independent Luttinger liquid, 2$n$ different velocities characterize the
ladder, meaning full mode separation. On the other hand, if the interband
scatterings are relevant, this coupling allows for $n$ different velocities
only. In the special case when the coupling strength depends on the relative
orientation of the spins of the scattered particles, the velocities of all
spin modes become equal, and the usual charge-spin separation is recovered.

Besides this weak-coupling fixed point there are several strong-coupling fixed
points. The relationship between them and the appropriate phases have been
determined by the bosonization treatment. The possible phases for generic
filling as well as for half-filled subbands and for a half-filled ladder have
been described. Since the exact location of the strong-coupling fixed points
cannot be determined in the approximation used, the strong-coupling regime has
been investigated numerically. The scaling equations have been solved for some
special initial values of the couplings, among others for a Hubbard ladder.

We have found that the generically filled Hubbard ladder has fully gapped
spin sector and one gapped charge mode \big(C1S0\big), the system does not
show Luttinger liquid behavior for any $n$. The half-filled Hubbard ladder is
not Luttinger liquid, either, it has not only fully gapped spin sector but the
charge sector, too, is gapped. The phase is C0S0. These results do not depend 
on the initial value of the Hubbard $U$, any nonzero value of the Coulomb 
interaction leads to the same phases.

The $n=2$ case, the SU(2) Hubbard ladder has been investigated in detailed
using similar methods in Refs.\ \onlinecite{LBMPAF-2} and \onlinecite{HHLLBMPAF}.
The results agree in that regime where the linearized spectrum is a reasonable
approximation. Six of the seven phases predicted in these papers were found by
us. The remaining one---as well as two of these six phases---are in that
parameter range where our approximation is questionable. Interestingly five of
these six phases occur in the phase diagram of the SU($n$) Hubbard ladder,
too.

The Hubbard ladder with half-filled subband shows special behavior in the case
$n=2$. We have found that in this case the backward scatterings are
irrelevant, therefore, there is one soft spin mode in the system (C0S1). On
the other hand if $n > 2$ the situation is the same as for a half-filled
ladder: all couplings are relevant and the corresponding phase is C0S0,
i.e. all modes are gapped. Similar result has been found in our earlier work,
for the SU($n$) symmetric half-filled Hubbard chain: the half-filled system
has gapped charge mode and soft spin mode (phase C0S1) for $n=2$, while for
$n>2$ the system is fully gapped (phase C0S0). In this sense similar behavior
has been found for the weakly coupled Hubbard ladder with half-filled subband
and for the half-filled Hubbard chain. This result seems to support the
claim\cite{KLH} that the intraband umklapp processes lead to
'confinement', i.e., to the vanishing of the interchain hopping in a
two-component weakly coupled two-leg Hubbard ladder.

Finally we have shown that if two $n$-component Luttinger chains with 
repulsive forward scattering are coupled by interchain hopping then---similarly 
to the case of the coupled generically filled $n$-component Hubbard chains---the
1D Luttinger liquid state is unstable not only for usual
electrons with spin-1/2 but for $n$-component fermions, too.
Luttinger liquid behavior could be obtained if some but not all of the couplings 
are attractive.

In the present paper, the gapped phases have been characterized by the 
number of gapped modes only. The study of the correlation functions
could reveal further details of the obtained phases. 

\begin{acknowledgments}
This research was supported in part by the Hungarian Research Fund (OTKA)
under Grant No. 43330. 
\end{acknowledgments}


\begin{thebibliography}{99}
  
\bibitem{FDMH} F. D. M. Haldane, J. Phys. C {\bf 14}, 2585 (1981).

\bibitem{EHLFYW} L. H. Lieb and F. Y. Wu, Phys. Rev. Lett. {\bf 20},
  1445 (1968).

\bibitem{JBMIA} J. B. Marston and I. Affleck, Phys. Rev. B {\bf 39},
  11538 (1989).

\bibitem{RAPAMCPL} R. Assaraf, P. Azaria, M. Caffarel, and P.
  Lecheminant, Phys. Rev. B {\bf 60}, 2299 (1999).

\bibitem{ESZJS} E. Szirmai and J. S\'olyom, Phys. Rev. B {\bf 71},
  205108 (2005).
  
\bibitem{CCCDCWM} C. Castellani, C. Di Castro and W. Metzner, Phys.
  Rev. Lett. {\bf 69}, 1703 (1992).

\bibitem{MF} M. Fabrizio, A. Parola, and E. Tosatti, Phys. Rev. B {\bf 46},
  3159 (1992); M. Fabrizio, A. Parola, Phys. Rev. Lett. {\bf 70}, 226 (1992);
  M. Fabrizio, Phys. Rev. B {\bf 48}, 15838 (1993).

\bibitem{DBCBAMST} D. Boies, C. Bourbonnais, and A.-M. S. Tremblay,
  Phys. Rev. Lett. {\bf 74}, 968 (1995); Proc. XXXI Rencontres de Moriond,
  Eds.: T. Martin, G. Montambaux, J. Tran Thanh (1996).

\bibitem{KLH} K. Le Hur, Phys. Rev. B {\bf 63}, 165110 (2001).

\bibitem{DVKTMR} D. V. Khveshchenko and T. M. Rice, Phys. Rev. B {\bf
    50}, 252 (1993).

\bibitem{HJS} H. J. Schulz, Phys. Rev. B {\bf 53}, R2959 (1996).

\bibitem{LBMPAF-2} L. Balents and M. P. A. Fisher, Phys. Rev. B
  {\bf 53}, 12133 (1996).

\bibitem{HHLLBMPAF} H.-H. Lin, L. Balents, and M. P. A.
  Fisher, Phys. Rev. B {\bf 58}, 1794 (1998).

\bibitem{CWWVLEF} C. Wu, W. V. Liu, and E. Fradkin, Phys. Rev. B
  {\bf 68} 115104 (2003).

\bibitem{GAJCNMH} G. Abramovici, J. C. Nickel, and M. H\'eritier,
  Phys. Rev. B {\bf 72}, 045120 (2005).

\bibitem{KPJS} K. Penc and J. S\'olyom, Phys. Rev. B {\bf 41}, 704
  (1990).

\bibitem{monte-carlo} E. Dagotto, J. Riera, and D. J. Scalapino, Phys.
  Rev. B {\bf 45} 5744 (1992).

\bibitem{monte-carlo-2} D. Poilblanc, H. Endres, F. Mila, M. G.
  Zacher, S. Capponi, and W.  Hanke, cond-mat/9605106.

\bibitem{DJS} D. J. Scalapino, J. Low Temp. Phys. {\bf 117}, 179 (1999).

\bibitem{RMNSRWDJS} R. M. Noack, S. R. White and D. J. Scalapino,
  Phys. Rev. Lett. {\bf 73}, 882 (1994).

\bibitem{YPSLTKL} Y. Park, S. Liang and T. K. Lee, Phys. Rev. B {\bf
    59}, 2587 (1999).

\bibitem{USSCJOFJBMMT} U. Schollw\"ock, S. Chakravarty, J. O.
  Fj{\ae}restad, J. B. Marston and M. Troyer, Phys. Rev. Lett. {\bf 90},
  186401 (2003).

\bibitem{AVR} A. V. Rozhkov, Phys. Rev. B {\bf 68}, 115108 (2003).

\bibitem{SCZ} S. C. Zhang, Science {\bf 275},1089 (1997).

\bibitem{DGSDS} D. G. Shelton and D. Sénéchal, cond-mat/9710251.

\bibitem{MTAF} M. Tsuchiizu and A. Furusaki, Phys. Rev. B {\bf 66},
  245106 (2002).

\bibitem{CHWH} C. Honerkamp and W. Hofstetter, Phys. Rev. Lett.
   {\bf 92}, 170403 (2004).

\bibitem{JS} J. S\'olyom, Adv. Phys. {\bf 28}, 201 (1979).

\bibitem{SLPKAF} S. Ledowski, P. Kopietz and A. Ferraz,
  cond-mat/0412620.



\end{thebibliography}
\end{document}